\pdfoutput=1
\documentclass[PAPER, cernpreprint, atlasdraft=false, UKenglish, texmf, orcidlogo]{atlasdoc}
\usepackage[utf8]{inputenc}
\usepackage{textcomp}
\usepackage{newunicodechar}
\newunicodechar{−}{\ensuremath{-}}
\usepackage[subcaption, full]{atlaspackage}
\usepackage{atlasbiblatex}

\usepackage{atlasphysics}
\graphicspath{{logos/}{figures/}}

\usepackage{ANA-EXOT-2025-09-PAPER-defs}
\usepackage{gensymb}
\AtlasTitle{Search for high-mass dilepton resonances in $pp$ collisions at \(\sqrt{s}=13.6\)~TeV combined with \(13\)~TeV results using the ATLAS detector}

\AtlasAbstract{
A search for spin-1 $ee$ and $\mu\mu$ resonances in the high-mass range is presented. Two proton-proton ($pp$) collision datasets recorded by the ATLAS detector at CERN's Large Hadron Collider are utilised. A search is first performed using a dataset recorded at a center-of-mass energy of $13.6$~TeV between 2022 and 2024, corresponding to an integrated luminosity of 165~fb$^{-1}$. The results of this search are subsequently statistically combined with results using 140~fb$^{-1}$ of data recorded between 2015 and 2018 at 13~TeV. The dilepton invariant mass spectrum is fitted with a functional form to model the contribution from Standard Model background processes. A generic spin-1 signal shape is used to determine the significance of observed deviations from the background expectation. No significant deviation is observed, therefore upper limits are placed at the 95\% confidence level on the fiducial cross-section times branching ratio for a spin-1 resonance with various width hypotheses, individually for $ee$, $\mu\mu$ final states and for their combination. From the combined 13~TeV and 13.6~TeV results, mass limits are extracted for various benchmark model bosons, including a $Z^{\prime}_{\text{SSM}}$ (5.5~TeV), $Z^{\prime}_{\chi}$ (5.1~TeV) and $Z^{\prime}_{\psi}$ (4.8~TeV). These results represent the most stringent limits to date on high-mass dilepton resonances on the considered $Z'$ models.
}

\AtlasRefCode{ANA-EXOT-2025-09}

\PreprintIdNumber{CERN-EP-2026-230}

\AtlasJournal{PLB}
\AtlasCoverEgroupAnalysisTeam{atlas-exot-2025-09-analysis-team@cern.ch}

\hypersetup{pdftitle={ATLAS document},pdfauthor={The ATLAS Collaboration}}

\begin{document}

\maketitle
\tableofcontents

\section{Introduction}

Significant discoveries have previously been made in searches in dilepton ($ee$ and $\mu\mu$) invariant mass distributions, including the discovery of the $J/\psi$ meson in 1974 ~\cite{JPsiDiscovery,JPsiDiscovery2,JpsiDiscoveryTing}, the $\Upsilon$ meson in 1977~\cite{UpsilonDiscovery}, and the $Z$ boson in 1983~\cite{SMZDiscovery,BAGNAIA1983130}. These channels provide an opportunity to both confirm the Standard Model (SM) and to search for physics beyond it. New models, including those introduced to address questions such as whether the strong and electroweak forces unify at high energy scales (GUTs)~\cite{GUTZPrimes}, or why the gravitational force is so weak (extra dimensional models)~\cite{Randall:1999ee, Arkani-Hamed:1998jmv},
predict deviations from the SM in the high-mass dilepton spectra.

This paper presents a search for high-mass resonances performed in a model-independent way. The search results are also interpreted for spin-1 vector boson resonances, generally referred to as $Z^{\prime}$, which often appear in models with extended gauge symmetries~\cite{GUTZPrimes}. Benchmark models include the Sequential Standard Model (SSM)~\cite{GUTZPrimes} in which the $Z^{\prime}$ has the same fermionic couplings as the SM $Z$ boson, and a resonance width divided by pole mass ($\Gamma/m_{\text{pole}}$, or relative width) of $\sim$3\%, and the more theoretically motivated GUT model~\cite{E6ZPrimes}, in which the $E_6$ gauge group is broken into SU(5) and two additional U(1) groups. The $Z^{\prime}$ candidate is composed of the lightest linear combination of the corresponding two new neutral gauge bosons, $Z^{\prime}_{\psi}$ and $Z^{\prime}_{\chi}$, such that \(Z^{\prime}(\theta_{E_{6}}) = Z^{\prime}_{\psi} \cos(\theta_{E_{6}}) + Z^{\prime}_{\chi} \sin(\theta_{E_{6}}) \), where \(0 \leq \theta_{E_{6}} < \pi \) is the mixing angle between the two gauge bosons.
The couplings to fermions is determined by the pattern of spontaneous symmetry breaking and the value of \(\theta_{E_{6}}\). In this search, resonances are assumed to have a narrow intrinsic width, comparable to the contribution from the detector mass resolution.
Two $E_6$ states are investigated as benchmarks, $Z^{\prime}_{\psi}$, the narrowest of the states with $\Gamma/m_{\text{pole}}\sim$~0.5\%, and $Z^{\prime}_{\chi}$, the broadest with $\Gamma/m_{\text{pole}}\sim$~1.2\%.

Experimentally, these models predict resonant local excesses of events over a smoothly falling dilepton mass spectrum.
The final states considered have the advantage of having a clean experimental signature with a fully reconstructed final state and excellent detection efficiency. Previous searches for exotic dilepton resonances have been performed in proton--proton collision data collected by ATLAS at $\sqrt{s}$ = 7~\TeV with total integrated luminosities of 40 pb$^{-1}$, $\sim$1~fb$^{-1}$ and 5~fb$^{-1}$~\cite{EXOT-2010-09, EXOT-2011-06, EXOT-2012-01}, at $\sqrt{s}$ = 8~\TeV using 20.3~fb$^{-1}$~\cite{EXOT-2012-23} and at $\sqrt{s}$ = 13~\TeV using 3.2~fb$^{-1}$, 36~fb$^{-1}$ and 139~fb$^{-1}$~\cite{EXOT-2015-07, EXOT-2016-05, EXOT-2018-08}.

Using the Run~2 data sample of 140~fb$^{-1}$ at $\sqrt{s}=13$~\TeV~\footnote{The quoted preliminary luminosity in the ATLAS Run~2 search was 139~fb$^{-1}$. The results have been rescaled to the newer and more precise luminosity measurement of 140~fb$^{-1}$.}, recorded in 2015--2018, and combining the $ee$ and $\mu\mu$ channels, the ATLAS and CMS collaborations have set lower mass limits on the $Z'_{\text{SSM}}$ ($Z'_{\psi}$) model at 95\% confidence level (CL) of 5.1 (4.5) and 5.2 (4.6)~\TeV, respectively~\cite{EXOT-2018-08, CMS-EXO-19-019}. ATLAS also set a mass limit of 4.8~\TeV on the $Z^{\prime}_{\mathrm{\chi}}$ model.

This search benefits two-fold over the ATLAS Run~2 result~\cite{EXOT-2018-08}. First, for the Run~3 data (recorded in 2022--2024 at a centre–of-mass energy of $\sqrt{s}=13.6$~\TeV), the $Z^{\prime}$ cross-sections are increased, due to the increase in centre-of-mass energy from 13 to 13.6~\TeV. For $m_{Z^{\prime}} = 6$~\TeV, $\sigma_{Z^{\prime}_{\mathrm{SSM}}}$ rises by 20\%. Second, combining the 2022--2024 Run~3 data sample (165~fb$^{-1}$) with the previous Run~2 only result (140~fb$^{-1}$), more than doubles the total data sample.

The results presented in this paper for resonant new physics in the dilepton invariant mass spectrum in Run~3 data follow the strategy and selection of the previous Run~2 result making adjustments for the Run~3 analysis where necessary. The results are then statistically combined with those from Run~2. The search uses a data-driven approach, fitting a functional form to the smoothly falling background spectrum in data to look for resonant excesses. The choice of fit function is selected by studies that optimise the fit to the background, which is estimated by using Monte Carlo (MC) simulation for most sources, and the matrix method~\cite{EXOT-2016-05} for the misidentified and non-prompt electron background. After performing the search, in the absence of a significant global excess, exclusion limits on the cross-section times branching ratio in both the $ee$ and $\mu\mu$ channels as well as the $ee$ + $\mu\mu$ combination are set for the 2022--2024 Run~3 and combined data samples.

\section{ATLAS detector}
\newcommand{\AtlasCoordFootnote}{%
ATLAS uses a right-handed coordinate system with its origin at the nominal interaction point (IP)
in the centre of the detector and the \(z\)-axis along the beam pipe.
The \(x\)-axis points from the IP to the centre of the LHC ring,
and the \(y\)-axis points upwards.
Polar coordinates \((r,\phi)\) are used in the transverse plane,
\(\phi\) being the azimuthal angle around the \(z\)-axis.
The pseudorapidity is defined in terms of the polar angle \(\theta\) as \(\eta = -\ln \tan(\theta/2)\) and is equal to the rapidity
$ y = \frac{1}{2} \ln \left( \frac{E + p_z}{E - p_z} \right) $ in the relativistic limit.
Angular distance is measured in units of \(\Delta R \equiv \sqrt{(\Delta y)^{2} + (\Delta\phi)^{2}}\).}

The ATLAS experiment~\cite{PERF-2007-01,GENR-2019-02} at the LHC is a multipurpose particle detector
with a forward--backward symmetric cylindrical geometry and a near \(4\pi\) coverage in
solid angle.\footnote{\AtlasCoordFootnote}
It consists of an inner tracking detector surrounded by a thin superconducting solenoid
providing a \qty{2}{\tesla} axial magnetic field, electromagnetic and hadronic calorimeters, and a muon spectrometer.
The inner tracking detector covers the pseudorapidity range \(|\eta| < 2.5\).
It consists of silicon pixel, silicon microstrip, and transition radiation tracking detectors.
Lead/liquid-argon (LAr) sampling calorimeters provide electromagnetic (EM) energy measurements
with high granularity within the region \(|\eta|< 3.2\).
A steel/scintillator-tile hadronic calorimeter covers the central pseudorapidity range (\(|\eta| < 1.7\)).
The endcap and forward regions are instrumented with LAr calorimeters
for EM and hadronic energy measurements up to \(|\eta| = 4.9\).
The muon spectrometer surrounds the calorimeters and is based on
three large superconducting air-core toroidal magnets with eight coils each.
The field integral of the toroids ranges between \num{2.0} and \qty{6.0}{\tesla\metre}
across most of the detector.
The muon spectrometer includes a system of precision tracking chambers up to \(|\eta| = 2.7\) and fast detectors for triggering up to \(|\eta| = 2.4\).
The luminosity is mainly measured by the LUCID--2 detector that is located close to the beampipe.
A two-level trigger system was used to select events~\cite{TRIG-2016-01,TRIG-2022-01}.
The first-level trigger is implemented in hardware and used a subset of the detector information
to accept events at a rate close to \qty{100}{\kHz}.
This is followed by a software-based trigger that
reduced the accepted rate of complete events to \qty{1.25}{\kHz} and \qty{3}{\kHz} on average in Run~2 and Run~3, respectively,
depending on the data-taking conditions.
A software suite~\cite{SOFT-2022-02} is used in data simulation, in the reconstruction
and analysis of real and simulated data, in detector operations, and in the trigger and data acquisition
systems of the experiment.

\section{Data and simulation}
\label{sec:dataSim}

The Run~2 data and simulation data samples are described in a previous paper~\cite{EXOT-2018-08}. The Run~3 data sample used in this analysis was collected during LHC stable beam conditions and with all detector systems operating normally. To ensure data quality, events with noise bursts or coherent noise in the calorimeters are removed. The integrated luminosity of the Run~2 and 2022--2024 Run~3 data samples are determined to be 140~fb$^{-1}$ and 165~fb$^{-1}$ respectively.
The uncertainty in the integrated luminosity for data recorded in 2015--2018 is 0.83\%~\cite{DAPR-2021-01} and in 2022--2024 is 1.9\%~\cite{ATL-DAPR-PUB-2025-001}, following the methodology discussed in Ref.~\cite{DAPR-2021-01}, using the LUCID-2 detector~\cite{LUCID2} for the primary luminosity measurements, complemented by measurements using the inner
detector and calorimeters.

The background contribution is estimated from data. However, simulated backgrounds and backgrounds estimated via data-driven techniques, described in Section~\ref{sec:modelling}, are used to find the optimal analytical fit function and assess the signal efficiencies and their uncertainties. The dominant background in this search is from the Drell--Yan (DY) process, which is irreducible since it has the same final state as the signal. Additional backgrounds estimated from MC simulation, listed from largest to smallest contribution, are: top-quark pair ($t\bar{t}$), single top-quark, and diboson production processes ($WW$, $WZ$ and $ZZ$).

Next-to-leading-order (NLO) in quantum chromodynamics (QCD) DY samples are used as the basis for both the background modelling and the construction of the signal templates.
The NLO DY samples are mass-sliced to ensure sufficient numbers of events throughout the dilepton invariant mass spectra.
Correction factors at next-to-next-to-leading order (NNLO) are recomputed for Run~3 using the same tool (\textsc{VRAP}~0.9~\cite{Anastasiou:2003ds}) and the same parton distribution functions (PDFs) as in the Run~2 analysis.

The spin-1 resonant signal templates are constructed from the NLO DY samples with the NNLO QCD corrections applied, using a signal reweighting method to enable a wide range of possible mass poles and widths to be created.
This method is the same as described in Ref.~\cite{EXOT-2016-05} but uses NLO DY instead of leading-order (LO) DY\@.
The signal reweighting procedure is based on the calculation of an event weight computed as the ratio of the signal plus DY cross-section to the background-only cross-section, where both models have the same initial and final states. This method rescales an SM background event to a desired signal $Z'$ model by choosing the appropriate couplings and charges of the signal model.
These templates are then used to determine the selection efficiency and the signal acceptance within the fiducial volume, defined at particle level to reduce the model dependence from off-shell effects, as in the Run~2 search.

The fiducial selection requires each electron or muon candidate to satisfy $|\eta| < 2.5$ and transverse momentum $p_{\mathrm{T}} > 30$~\GeV, and the dilepton invariant mass to satisfy $m_{\ell\ell} > m_{\text{pole}} - 2\Gamma_{X}$, where $m_{\text{pole}}$ and $\Gamma_{X}$ represent the pole mass and width of the hypothetical resonance $X$, respectively. Interference effects between the resonant signal and the background processes are neglected.

Electroweak (EW) corrections at NLO were also applied to the DY sample to account for any remaining EW-scheme-dependent effects. These corrections have the same values as in Run~2 (calculated by \textsc{MCSANC}~\cite{Arbuzov2016}) due to their stability against changes in the centre-of-mass energy.
The DY sample with both NNLO QCD and NLO EW corrections is used to perform the background modelling studies and to obtain detector resolution functions as a function of the invariant mass. The latter enables generic signal models to be built, which are used in the signal modelling described in Section~\ref{sec:modelling}.

Several dedicated $Z^{\prime}$ ($E_6$ and SSM model) LO simulated signal samples were generated to validate the signal reweighting technique and resulting templates. The theoretical fiducial NNLO cross-sections for each model were computed as the product of the LO cross-section, an LO-to-NNLO QCD correction factor, and the acceptance determined using the signal-reweighted DY samples. The higher-order LO-to-NNLO QCD correction factor was recomputed for 13.6~\TeV using the same methodology and PDFs as in the Run~2 search.

For all samples the detector response was simulated with \GEANT~\cite{Agostinelli:2002hh} and the events were processed with the same reconstruction software as used for the data. The distribution of the number of additional simulated proton--proton collisions in the same or neighbouring beam crossings (pile-up) is accounted for by overlaying simulated minimum-bias events and re-weighting the simulated events to match the distribution observed in the data. The event generators for the hard-scatter matrix element, as well as showering, and PDFs that were used are listed in Table~\ref{Tbl:Bkgds}.

\begin{table}[h]
\caption{Summary of generators used for simulation of background and signal samples. `Tune' refers to the set of tuned parameters used in the simulation.}
\label{Tbl:Bkgds}
\footnotesize
\centering
\resizebox{\linewidth}{!}{
\begin{tabular}{llllll}
\toprule
Physics process   & Generator & Parton shower &  Normalisation & Tune & PDF (generator) \\\midrule
Drell--Yan & \POWHEGBOX[v2]~\cite{Frixione:2007nw,Nason:2004rx,Frixione:2007vw,Alioli:2010xd} & \PYTHIA[8.312]~\cite{Sjostrand:2014zea} & NLO & AZNLO~\cite{STDM-2012-23} &  CT10~\cite{Lai:2010vv}\\
\ttbar, Single-top & \POWHEGBOX[v2]  & \PYTHIA[8.308] & NLO & A14~\cite{ATL-PHYS-PUB-2014-021} & NNPDF2.3lo~\cite{Ball:2012cx}  \\
Diboson & \SHERPA[2.2.14]~\cite{Bothmann:2019yzt} & \SHERPA[2.2.14] & LO--NLO & Default& NNPDF3.0nnlo~\cite{Ball:2014uwa}\\
$Z^{\prime} (\text{SSM}, E_6)$ & \PYTHIA[8.312] & \PYTHIA[8.312] & LO & A14 & NNPDF2.3lo \\
\bottomrule
\end{tabular}
}

\end{table}

\section{Event selection}

For the $ee$ channel, events are required to have been selected by an unprescaled trigger requiring two electrons both satisfying a 24~\GeV $p_{\text{T}}$ threshold. In the $\mu\mu$ channel, events are required to have satisfied one of two unprescaled single muon triggers, with $p_{\text{T}}$ thresholds of 24~\GeV and 50~\GeV.
The 24~\GeV muon trigger also used a $p_{\text{T}}$-dependent isolation threshold to maximise the selection efficiency~\cite{TRIG-2022-01}. Selected events are required to satisfy basic cleaning criterion designed to remove spurious events from instrumental noise.  The remaining events are then required to contain at least two good quality electrons or muons.

Electron candidates are considered good quality if they are within $|\eta| = $ 2.47, excluding the calorimeter transition region 1.37 $< |\eta| <$ 1.52, whereas muons are required to be within $|\eta| = $ 2.5, and both electrons and muons are required to have $p_{\text{T}}>$ 30~\GeV. Electrons (muons) are also required to satisfy track-based requirements such as being associated with a primary vertex, having a transverse impact parameter $d_0$ significance less than five (three), and $|z_0\sin{\theta}|$ less than 0.5 (0.5)~mm, where the $z_0$ is longitudinal impact parameter and $\theta$ is the polar angle of the track. Electrons are further required to be isolated within the calorimeter using a fixed-sized cone designed to have a high background rejection at high $p_{\text{T}}$, while muons are required to be isolated from tracks within a tight $p_{\text{T}}$-dependent variable-sized cone~\cite{MUON-2018-03}. Electrons are required to satisfy `tight' identification requirements to suppress sources of fake background contributions~\cite{EGAM-2021-01,EGAM-2018-01, PERF-2017-01}, and muons are required to satisfy `high-$p_{\text{T}}$' requirements to ensure the highest resolution and quality muons, even in the multi-\TeV range~\cite{MUON-2022-01,MUON-2018-03}. Furthermore, corrections are applied to simulated electron (muon) candidates to match the energy (momentum) scale, resolution and efficiencies of the simulation to the data~\cite{EGAM-2021-01, PERF-2017-03, EGAM-2018-01, PERF-2017-01, MUON-2022-01, MUON-2018-03}.

From all of the electron and muon candidates that satisfy the selection criteria, pairs of same flavour leptons are formed. In the case of multiple possible dilepton pair candidates in the event, the two highest $p_{\text{T}}$ same-flavour leptons are used to form the pair. If both an $ee$ and $\mu\mu$ pair are selected within the same event, the $\mu\mu$ candidate is rejected in favour of the higher efficiency and better resolution of the $ee$ channel. The selected muons within the $\mu\mu$ pair must be oppositely charged, whereas the electrons within the $ee$ pair do not have a charge requirement imposed because of the higher probability of charge misidentification for high-$p_{\text{T}}$ electrons. Finally the $ee$ and $\mu\mu$ invariant mass is required to be $>130$~\GeV to ensure that the selected events are above the $Z$ boson mass peak.

\section{Signal and background modelling}
\label{sec:modelling}

The background contribution is determined using the observed $ee$ and $\mu\mu$ invariant mass spectra, to which smooth background and model-independent signal shape components are fitted. The background fit function is determined from a family of smoothly falling functions, following the procedure used in Run 2.
These potential functions are tested against the expected background contributions described in Section~\ref{sec:dataSim} and, for the $ee$ channel, additional background contributions from multijet and $W$+jets, referred to as `fakes'. In such cases, one or both electron candidates come from mis-reconstruction of hadronic jets or non-prompt hadron decays. These background contributions are estimated by using the data-driven matrix method. The contributions from leptonically decaying $\tau$-leptons in both channels, and the multijet, $W$+jets processes and non-prompt contributions in the $\mu\mu$ channel are assessed and found to be negligible. The signal shapes are obtained from the convolution of non-relativistic Breit--Wigner functions of various widths with the detector resolution~\cite{BW1,BW2}.  The parameterisation of the detector resolution is validated using the zero-width generic signals. Several fit tests are used to optimise the functional form choice, as was done in Run 2 and described in more detail below.

The functional form of the background estimate, used in the fit to data, is given by:
\begin{equation} \label{eq:bkg_function}
f\left(x \equiv \frac{m}{\sqrt{s}}\right) = f_{\text{BW},Z}(m) \times (1 - x^\alpha)^{p_1} x^{-\sum_{i=2}^N p_i (\ln x)^{i-2}} \, ,
\end{equation}
where the function $f_{\text{BW},Z}(m)$ is a non-relativistic Breit-Wigner function with $m_{Z}$~=~91.2~\GeV and $\Gamma_{Z}$~=~2.5~\GeV~\cite{Zyla:2020zbs}, $m$ is the dilepton invariant mass, $\sqrt{s}$ the centre-of-mass energy, $N$ the number of free parameters $p_i$ that describe the shape, and $\alpha$ sets the exponent of the variable $x$.
The two chosen parameters in this function are $\alpha$ and $N$, which are determined based on spurious signal tests, signal injection tests and F-tests, as was done in Run 2 and are described in more detail below.

First, background-only fits to the total background are performed in the $ee$ ($\mu\mu$) channel in 1~\GeV bins over the range 0.5--7.1 (7.6)~\TeV in invariant mass, for degrees of freedom (given by $N$ in Eq.~(\ref{eq:bkg_function})) between four and seven. The mass ranges differ in the two channels to reduce edge effects due to the different resolutions of the two dilepton channels, resulting in the search pole mass range from 0.58 to 6.1~\TeV in both channels.
Fits are required to have a global fit $\chi^{2}$-probability $>$ 0.05 to eliminate functions that cannot describe the basic background shape.
A function with $N$ degrees of freedom is tested by fitting Asimov data samples created from fits using functions with $N+1$ degrees of freedom.

Second, spurious signal tests are performed for these background-only Asimov data samples. These tests consist of signal plus background (S+B) fits, and allow functions that produce fake (spurious) signals or show instability across templates to be eliminated from further consideration. The spurious signal ($N_\text{sig}$) and its uncertainty ($\sigma_\text{fit}$) are obtained from the resulting signal yield and its uncertainty after the fit to the background-only Asimov distribution. Fits satisfy the spurious signal requirement if $|N_\text{sig}/\sigma_\text{fit}|$ $<$ 0.5. This calculation is repeated for different signal widths (0, 0.5\%, 1.2\%, 3\%, 6\%, 10\%). The final step is to select one function, from those that satisfy the spurious signal requirement, using the F-test with Asimov and pseudo-data samples. The F-test evaluates how increasing the number of degrees of freedom in the function affects the background fit quality.
The functional form with the lowest number of degrees of freedom that describes the background well and has the smallest absolute number of fitted signal events (spurious signal) across the whole $ee$ and $\mu\mu$ invariant mass range is selected~\cite{ATL-PHYS-PUB-2020-028}. Signal injection tests are also performed to ensure that the correct signal yields can be extracted from the S+B fit.

For Run~2, the $ee$ channel used $\alpha = 1$ and $N = 5$, and the $\mu\mu$ channel used $\alpha = \frac{1}{3}$ and $N = 5$, while for Run~3, both channels use $\alpha = 1$ and $N = 5$. The difference between the $\alpha$ parameters is mainly due to the different centre-of-mass energy of the datasets, but also to the different fit range used in both searches.

\section{Systematic uncertainties}

The results of this search are generally dominated by statistical uncertainties given the high-mass target; however, there are also important systematic uncertainties that are considered. The background modelling uncertainty arises from the choice of the analytic function used to describe the smoothly falling dilepton mass spectrum. A spurious signal uncertainty is derived from S+B fits to background-only templates, including variations from the electron fake background estimate, DY theory uncertainties, and the higher-order NNLO QCD corrections. This uncertainty is parameterised as a function of the resonance mass and width, and is included as another contribution to the total yield -- with its own nuisance parameter -- having the same shape as the signal being evaluated.

Systematic uncertainties affecting the signal acceptance and detector response are also evaluated. In the $\mu\mu$ channel, the dominant uncertainties
arise from reconstruction (including identification), momentum resolution and the `bad-muon' veto (which helps to reduce the impact of `pathological' tracks)~\cite{MUON-2022-01}. In the $ee$ channel, the dominant systematic uncertainties arise from electron energy resolution~\cite{EGAM-2021-02} and isolation~\cite{EGAM-2021-01}.

The signal PDF uncertainties are estimated by using the nominal PDF set described in Table~\ref{Tbl:Bkgds} and its 100 PDF eigenvector variations.
Uncertainties due to $\alphas$ are estimated by varying $\alphas(m_Z)$ by $\pm 0.002$. The impact of missing higher order corrections is evaluated using seven-point variations of the factorisation and renormalisation scales in the cross-section calculations. Both scales are varied upwards and downwards by a factor of two, considering only same-direction variations. The envelope of the resulting variations as a function of $m_{\ell\ell}$ is taken to represent the associated systematic uncertainty. The final signal modelling uncertainty is computed as the sum in quadrature of the PDF, $\alphas$ and individual scale uncertainties.

Experimental uncertainties are treated as uncorrelated between the Run~2 and Run~3 data samples, while modelling uncertainties in the signal are correlated across the two periods. A summary of the systematic uncertainties considered and their impacts are shown in Table~\ref{tab:systematics:summary}.
The values shown in the table are computed as the impact on the signal yield for those uncertainties affecting the mentioned quantity, and as the change in the mean and width of the parameterisation of the detector resolution for the energy scale and resolution uncertainties, respectively.
Signal modelling uncertainties are also computed as the change in the signal yield.
The values of the spurious signal uncertainty shown in the table are computed as the fraction of the total spurious signal yield to the predicted cross-section for a 3\%-width $Z'_{\text{SSM}}$ signal at the given mass, in percentage.

\begin{table}[ht!]
\centering
\caption{Impact of the various systematic uncertainties considered in the Run~3 analysis. These are shown as percentages and computed with zero-width signals for uncertainties other than the spurious signal, which is computed with 3\%-width signals. The impacts of uncertainties affecting the signal yield are calculated as the change in that quantity, while impacts on the energy scale and resolution are computed as the change in the detector resolution parameterisation. The spurious signal impact shown is computed as the fraction of the total spurious signal to the predicted cross-section of a $Z'_{\text{SSM}}$ signal.
In the $ee$ channel, reconstruction uncertainties also comprise identification systematic uncertainties. The `--' denotes where a given systematic uncertainty is not applicable.}
\begin{tabular}{|l|S[table-format=<2.1]S[table-format=<2.1]|S[table-format=<2.1]S[table-format=<2.1]|}
\hline
Systematic & \multicolumn{2}{c|}{$ee$ channel [\%]} & \multicolumn{2}{c|}{$\mu\mu$ channel [\%]} \\
\hline
Mass [\TeV] & {2} & {6} & {2} & {6} \\
\hline
Spurious signal       & <1 & <1 & <1 & 37 \\
\hline
Signal modelling       & 4.5 & 23 & 4.5 & 23 \\
\hline
\textit{Applied to signal yield:}       & \multicolumn{4}{c|}{} \\
Reconstruction                                          & 4.3 & 4.3 & 10 & 55 \\
Isolation                                               & 7.3 & 18 & 1.7 & 1.8 \\
Trigger                                                & <1 & <1 & 8.4 & 9.3 \\
Bad muon veto                                                & {--} & {--} & <1 & 52 \\
\textit{Applied to signal shape:}       & \multicolumn{4}{c|}{} \\
Energy/momentum scale                              & <1 & 1.0 & <1 & <1 \\
Energy/momentum resolution            & 54 & 61 & 20 & 16 \\
\hline
\end{tabular}
\label{tab:systematics:summary}
\end{table}

\section{Statistical interpretation}

Binned profile likelihood fits are used to interpret the agreement between the data, background prediction, and various possible signal hypotheses taking into account the systematic uncertainties described above. Those uncertainties are accounted for as log-normal (in the case of the luminosity uncertainty), or Gaussian (for all other uncertainties) distributed nuisance parameters. The likelihood fits are performed using the \mll~spectra, in the range of 0.5~\TeV to 7.1 (7.6)~\TeV using 1~\GeV bins, separately for the $ee$ ($\mu\mu$) channel. For the combined dilepton channel the range 0.5--7.1~\TeV is used with 1~\GeV bin widths.
Fits are performed for Run~3 only, and combined Run~2 and Run~3 data samples. Using Poisson statistics, the log likelihood expression can be written as:

\begin{equation}
\mathcal{L}(n|s,b,\theta) =
\prod_{i=1}^{N_{\text{channel}}} \prod_{j=1}^{N_{\text{bins}}} \text{Pois}\left(n_{ij} | \mu s_{ij} + b_{ij}\right)\prod_{k=1}^{N_{\text{sys}}}G(\theta_k),
\end{equation}

where \(n_{ij}\) is the number of measured data events in channel \(i\) and mass bin \(j\), with the same indices for expected signal \(s\) and background \(b\) counts, estimated from MC simulation and the functional fit, respectively. The nuisance parameters are denoted by $\theta$, which can take various distributed forms \(G\) as appropriate. The parameter \(\mu\) is referred to as the signal strength and is a common normalisation factor to the signal yield in all bins. One of the goals of the fit is the determination of this parameter, also known as the parameter of interest (PoI).

The significance of an excess is quantified based on the probability that the background fluctuations create
an excess with the same characteristics as the signal equal to or larger than the one observed. The probability \pval of a local excess, $p_{0}$, is determined from a profile-likelihood-ratio test statistic using the asymptotic approximation~\cite{Cowan:2010js}. The `look-elsewhere' effect accounting for the trial factors associated with searching for multiple signal hypotheses in the same data sample is covered by the global significance calculation~\cite{Gross_2010}. Global significances for fluctuations with local significances above two standard deviations ($\sigma$) can be calculated using the asymptotic method:

\begin{equation*}
p_\text{global} = p_\text{local} + N_\text{up} \cdot \exp\left(-\frac{1}{2}Z_\text{local}^2\right).
\end{equation*}

In this equation, $N_\text{up}$ is the number of down-crossings of the $p$-value distribution at 0.5, $p_\text{local}$ denotes the $p$-value of the most significant observed excess, while $Z_\text{local}$ is the corresponding significance, and $p_\text{global}$ the resulting global significance. In this search a range of pole masses are considered for separate width hypotheses, with local and global significances quoted for each width hypothesis.

In the absence of any significant excess, upper exclusion limits are set at 95\% CL on the fiducial cross-section times branching ratio in the $ee$, $\mu\mu$ and combined dilepton channels. The limits are evaluated with the modified frequentist CL$_{\mathrm{S}}$ method using the asymptotic approximation~\cite{Read:2002hq}.

For the combination of results from Run~2 and Run~3, the separate initial likelihoods are multiplied together, creating an inherent model dependence on the signal strength parameter, given the signal cross-section difference between the 13 and 13.6~\TeV centre-of-mass energies. For this reason, the fully combined result is only quoted for the various benchmark models and not the generic model-independent signal hypotheses.

\section{Results}

\begin{figure}
\centering
\begin{subfigure}[b]{0.48\textwidth}
\centering
\includegraphics[width=\textwidth]{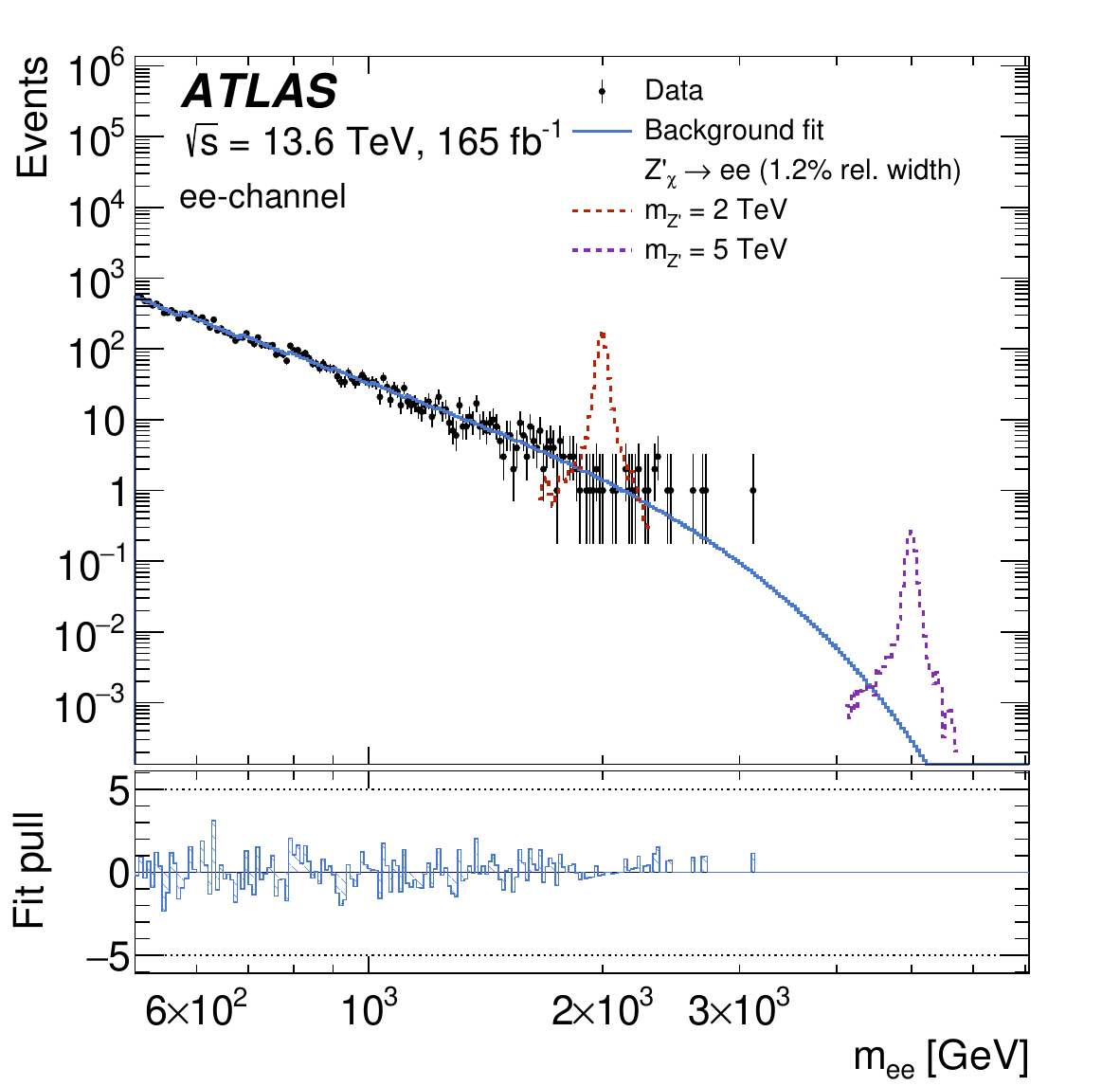}
\caption{}
\label{fig:m_ee}
\end{subfigure}
\hfill
\begin{subfigure}[b]{0.48\textwidth}
\centering
\includegraphics[width=\textwidth]{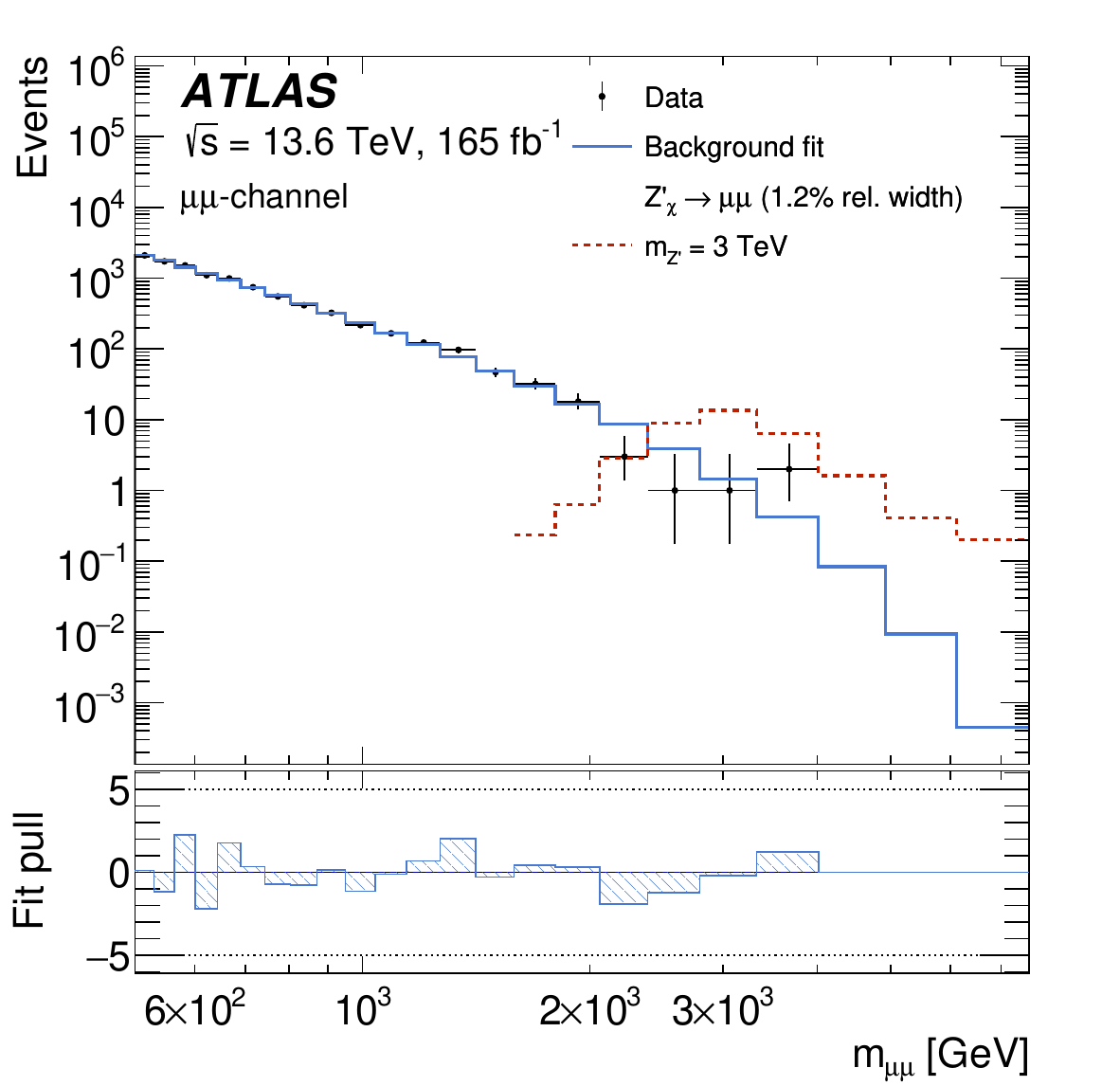}
\caption{}
\label{fig:m_mm}
\end{subfigure}
\caption{The invariant mass spectrum for the (a) $ee$ and (b) $\mu\mu$ channels from the Run~3 data. The observed data and statistical uncertainties are shown in the markers and vertical error bars, while the blue histogram shows the background fit. $Z'_{\chi}$ signal models with varying masses and relative width of 1.2\% are overlaid and represented with dashed lines, each normalised to its expected yield according to the predicted cross-section.
The bottom panels show the fit residuals normalised by the statistical uncertainty in the data (pull) at each bin.
The data samples are displayed with increasing bin widths starting from 6~\GeV (30~\GeV) in the $ee$ ($\mu\mu$) channel, according to their respective detector resolutions.
}
\end{figure}

The Run~3 invariant mass distributions for the $ee$ and $\mu\mu$ events that satisfy the full analysis selection, and the background fit are displayed in Figure~\ref{fig:m_ee} and \ref{fig:m_mm}, respectively, with varying bin widths dependent on mass resolution.
The highest invariant mass event observed in the $ee$ channel is 3.13~\TeV and in the $\mu\mu$ channel is 3.96~\TeV. The leading electron in the former event has $p_{\text{T}} =$ 883~\GeV, $\eta =$ 1.15 and $\phi = -1.55$ and the sub-leading electron has $p_{\text{T}} =$ 534~\GeV, $\eta = -1.87$ and $\phi =$ 0.17. The leading muon in the latter event has $p_{\text{T}} =$ 3769~\GeV, $\eta = -0.19$ and $\phi = -0.18$ and sub-leading muon has $p_{\text{T}} =$ 937~\GeV, $\eta =$ 0.17 and $\phi =$ 2.97. The second event with highest invariant mass in the $\mu\mu$ channel has a mass of 3.6~\TeV, with the leading muon having $p_{\text{T}} =$ 574~\GeV, $\eta =$ 1.56 and $\phi = 2.94$ and the sub-leading muon having $p_{\text{T}} =$ 527~\GeV, $\eta = -2.13$ and $\phi = -0.12$.
This Run~3 highest invariant mass event surpasses that observed in Run~2 in the $\mu\mu$ channel  (2.75~\TeV), but not in the $ee$ channel (4.06~\TeV).

\begin{figure}
\centering
\begin{subfigure}[b]{0.49\textwidth}
\centering
\includegraphics[width=\textwidth]{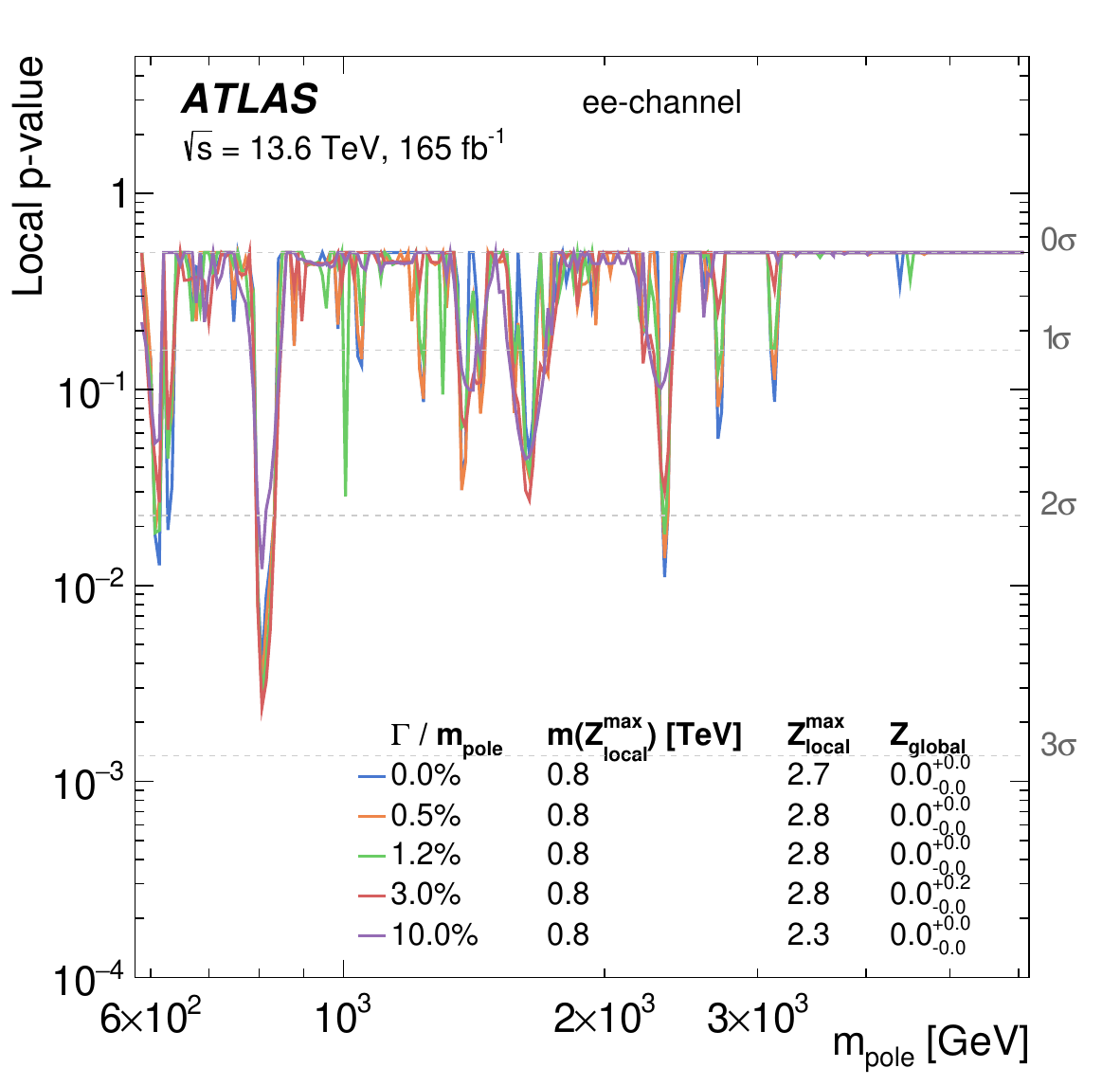}
\caption{}
\label{fig:pvals_ee}
\end{subfigure}
\hfill
\begin{subfigure}[b]{0.49\textwidth}
\centering
\includegraphics[width=\textwidth]{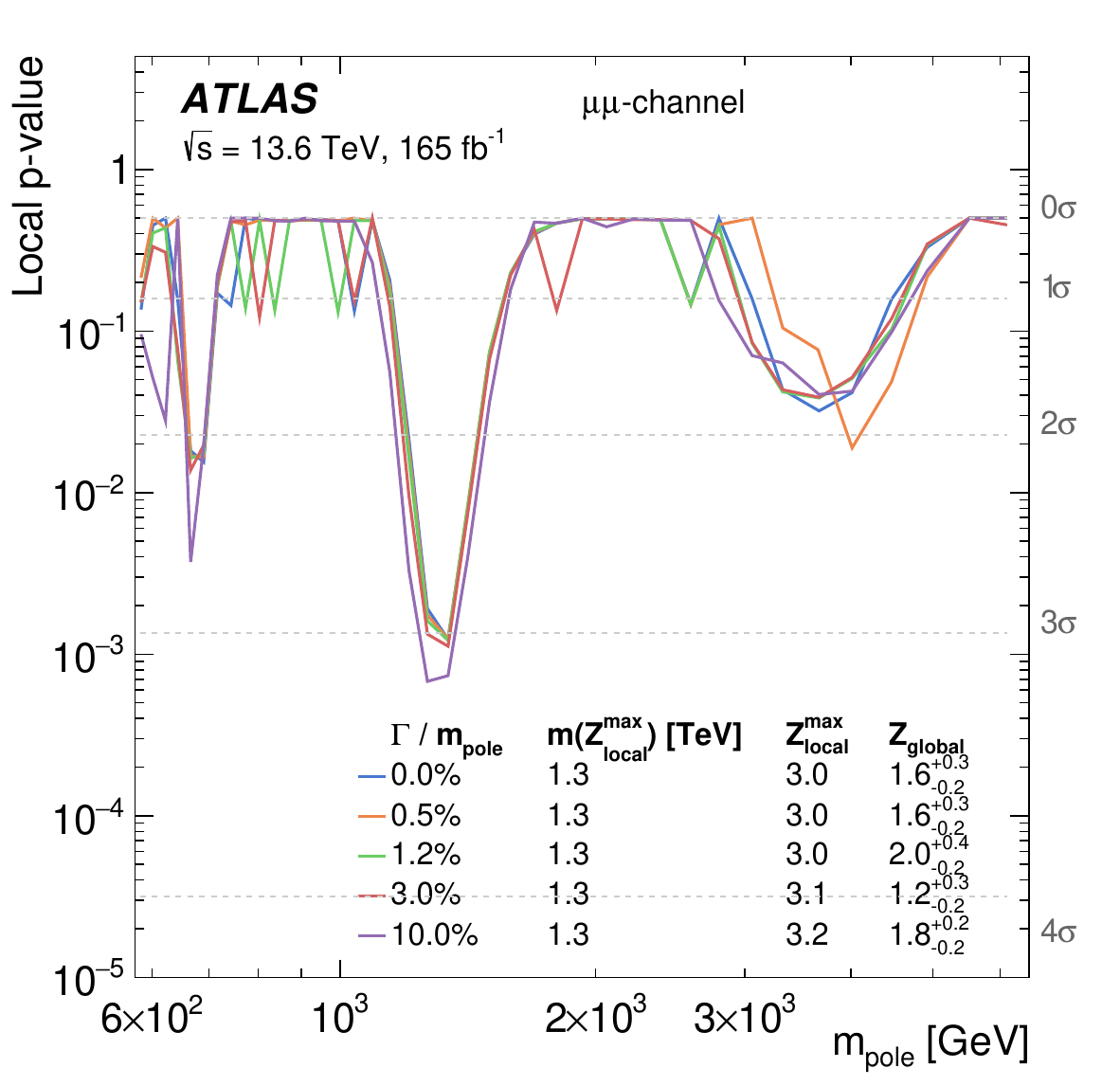}
\caption{}
\label{fig:pvals_mm}
\end{subfigure}
\hfill
\begin{subfigure}[b]{0.6\textwidth}
\centering
\includegraphics[width=\textwidth]{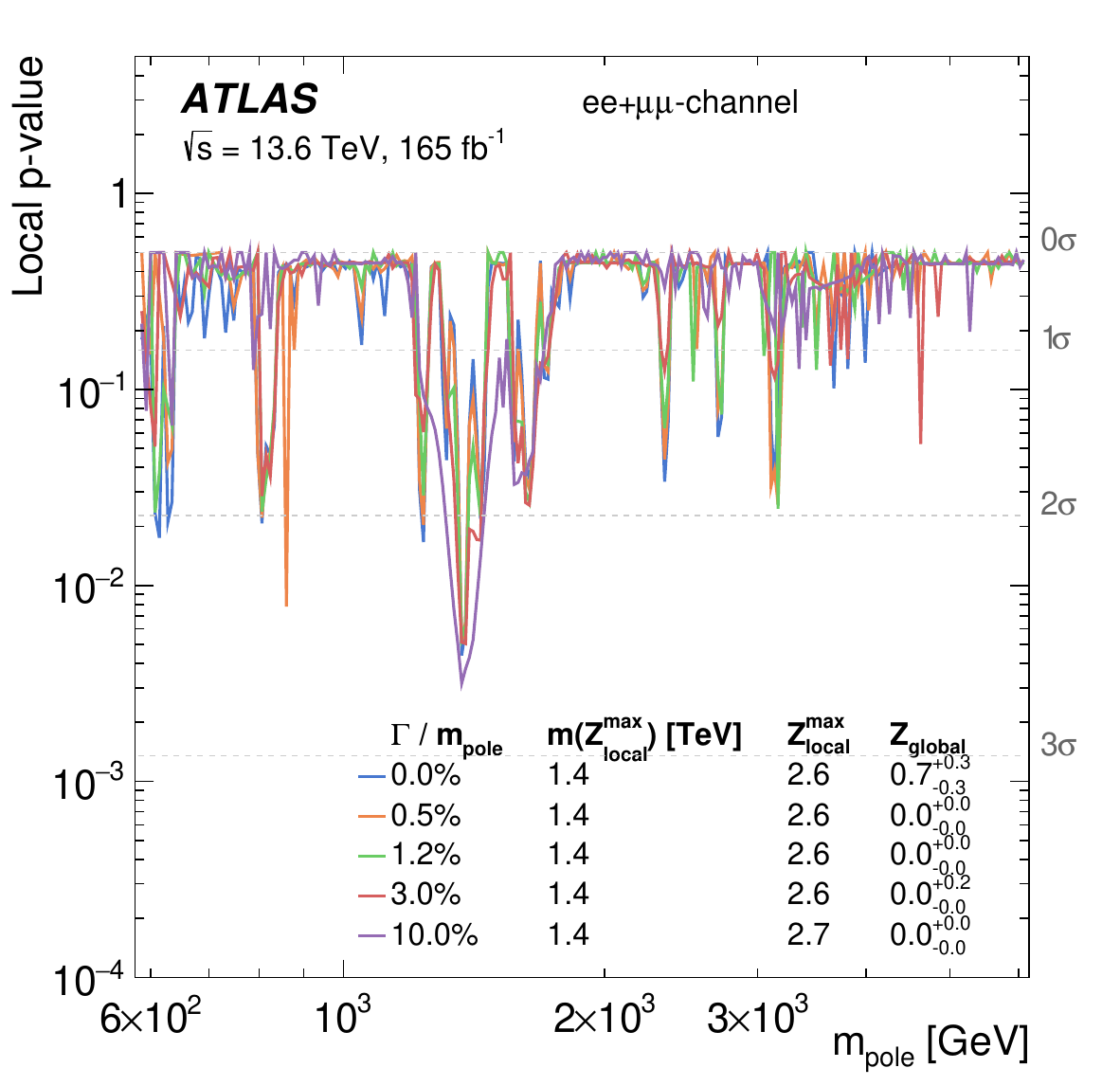}
\caption{}
\label{fig:pvals_combined}
\end{subfigure}
\caption{The probability that the observed mass spectrum is compatible with the background-only hypothesis ($p$-value) for the (a) $ee$, (b) $\mu\mu$ and (c) combined dilepton channels. The local $p_{0}$ is quantified in standard deviations $\sigma$ as a function of pole mass $m_{\text{pole}}$. The maximum local significance $Z^{\mathrm{max}}_{\mathrm{local}}$ is the largest local significance over the tested mass range, and $Z_{\mathrm{global}}$ is the corresponding global significance.}
\label{fig:pvals}
\end{figure}

The fit to data to evaluate the background-only hypothesis, described in Section~\ref{sec:modelling}, is performed in bins of 1~\GeV in the dilepton invariant mass range for the $ee$ ($\mu\mu$) channel from 0.5 to 7.1 (7.6)~\TeV. Signal pole masses are tested, and $p$-values calculated in both channels ranging from 0.58 to 6.1~\TeV, with spacing dictated by the detector resolution.
The probability that the  data are compatible with the background-only hypothesis is shown in Figure~\ref{fig:pvals} as a function of pole mass, for various width hypotheses. Figure~\ref{fig:pvals_ee}, \ref{fig:pvals_mm} and \ref{fig:pvals_combined} show the local $p$-values for $ee$, $\mu\mu$ and combined dilepton channels with five varying $Z^{\prime}$ relative widths: %
0\%, 0.5\%, 1.2\%, 3.0\% and 10\%.

For the $ee$ channel, the largest local excess, of 2.8$\sigma$, occurs at 0.8~\TeV for all widths, however, the global significances are consistent with zero. The $\mu\mu$ channel has a maximum 3.2$\sigma$ local excess at 1.3~\TeV for all widths, with a width-dependent global significance ranging from 1.2 to 2.0$\sigma$. The combined dilepton data samples indicate a 2.6$\sigma$ deviation for all width hypothesis at 1.4~\TeV, with a global significance consistent with zero for all except zero width (0.7$\sigma$).
Therefore, no globally significant deviation is observed for all channels and widths tested.

\begin{figure}
\centering
\begin{subfigure}[b]{0.49\textwidth}
\centering
\includegraphics[width=\textwidth]{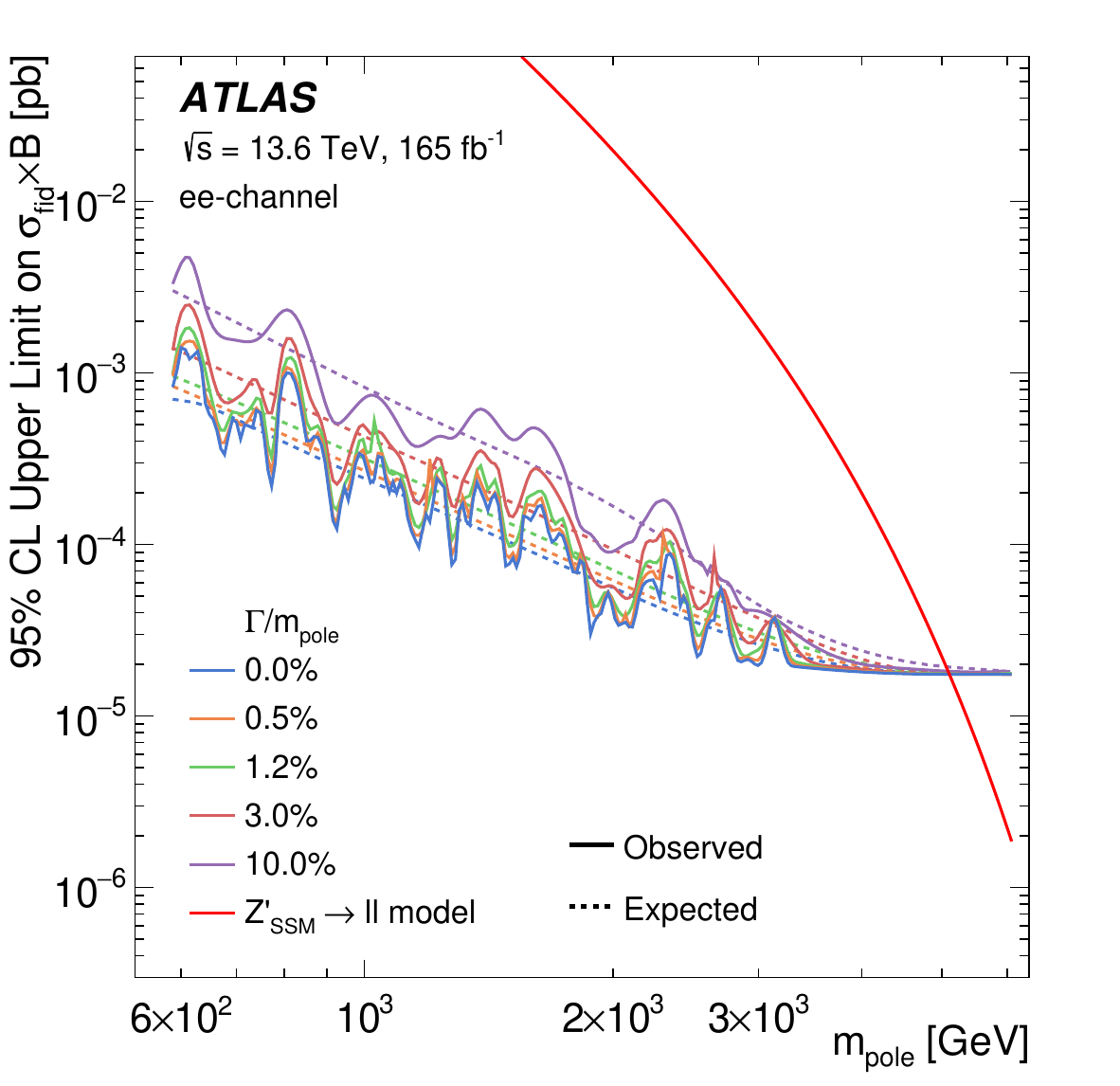}
\caption{}
\label{fig:width_limits_ee}
\end{subfigure}
\hfill
\begin{subfigure}[b]{0.49\textwidth}
\centering
\includegraphics[width=\textwidth]{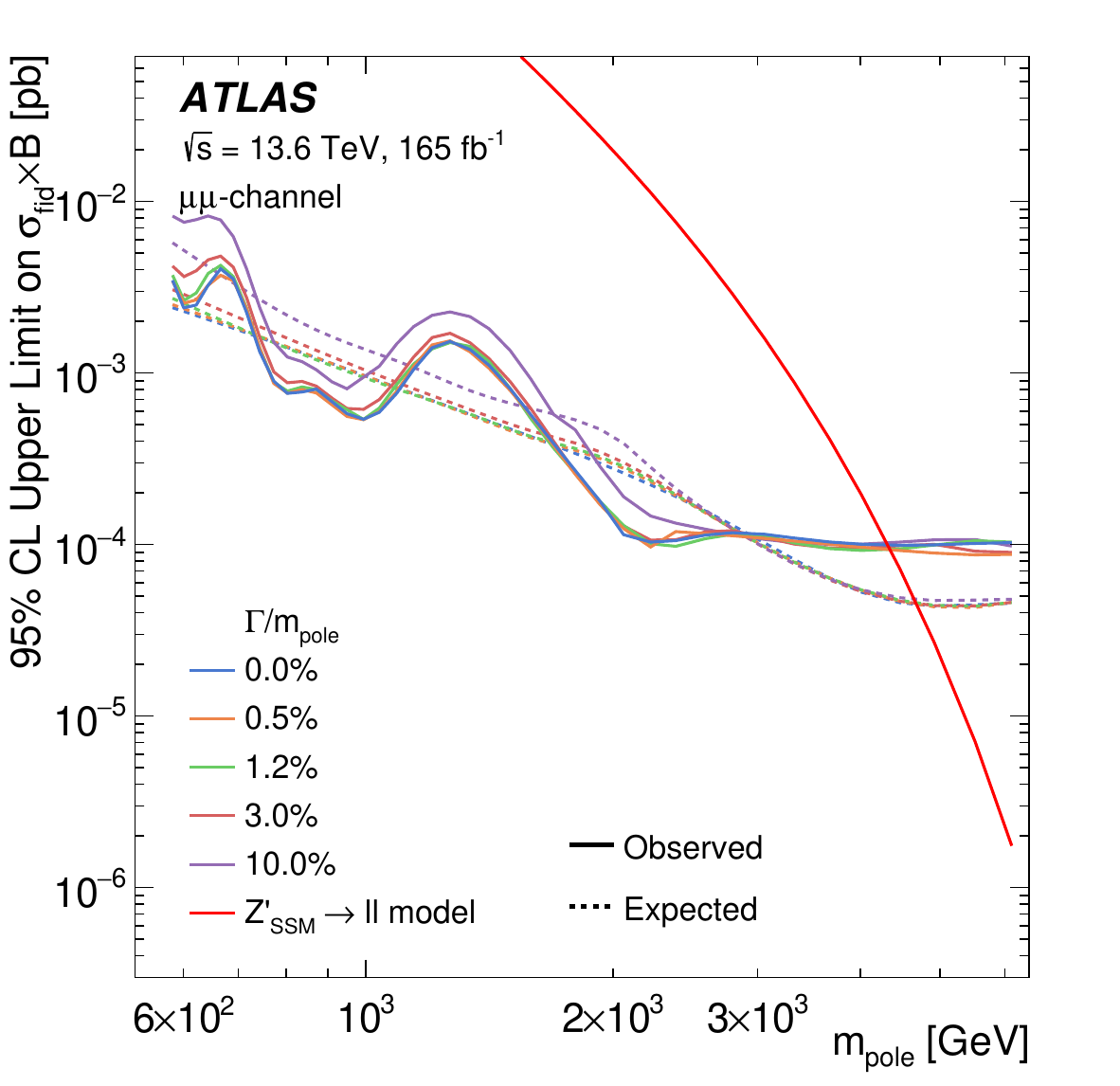}
\caption{}
\label{fig:width_limits_mm}
\end{subfigure}
\hfill
\begin{subfigure}[b]{0.6\textwidth}
\centering
\includegraphics[width=\textwidth]{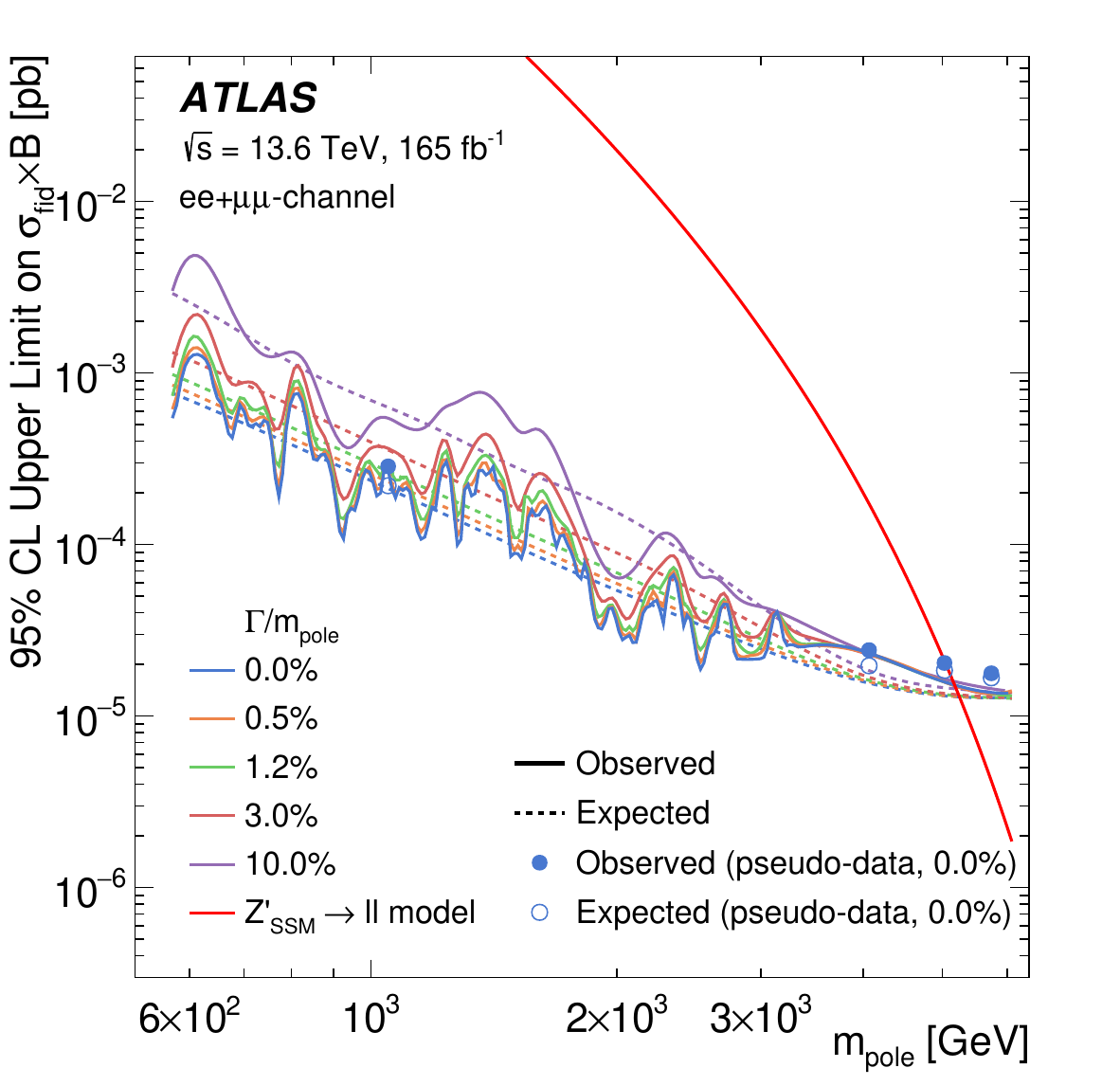}
\caption{}
\label{fig:width_limits_combined}
\end{subfigure}
\caption{Observed and expected upper limits at 95\% CL on the fiducial cross-section times branching ratio ($\sigma_{\text{fid}} \times \text{B}$) as a function of pole mass for the (a) $ee$, (b) $\mu\mu$ and (c) combined dilepton channels, for zero-width and 0.5\%, 1.2\%, 3\% and 10\% relative width signals. Observed and expected limits are indicated by the solid line and dashed line, respectively. Superposed are the theoretical cross-sections times the single-flavour-lepton pair branching ratio for $Z^{\prime}_{\mathrm{SSM}}$ ($\Gamma$/$m_{\text{pole}}$ = 3.0\%) in the fiducial region.
Results of expected and observed limits for zero-width signals computed using pseudo-data are overlaid for four $m_{\text{pole}}$ points in the combined dilepton channel figure as empty and filled markers, respectively.
}
\end{figure}

Exclusion limits are computed using signals with mass spacing to match the experimental resolution, and are then smoothed~\cite{dierckx1975algorithm,dierckx1982fast,dierckx1981improved,dierckx1995curve}.
The upper exclusion limits, at 95\% CL, on the fiducial cross-section times branching ratio to a single-flavour lepton pair as a function of pole mass for various relative signal widths are shown in Figure~\ref{fig:width_limits_ee}, \ref{fig:width_limits_mm}, and \ref{fig:width_limits_combined} for the $ee$, $\mu\mu$ and combined dilepton channels, respectively. Lower mass limits are then extracted from the intersection of this curve for a particular width with the theoretical cross-section curve. The results of observed and expected limits computed using pseudo-data for zero-width signals at four masses are also shown in the Figure~\ref{fig:width_limits_combined}.
Running with sufficient pseudo-data for all mass points including systematic uncertainties is computationally prohibitive, however, to compare how aggressive using the asymptotic approximation is, limits using both approaches are compared at various mass points.
The difference between fiducial cross-section limits obtained using the asymptotic approximation rather than pseudo data increases with the centre-of-mass. The fiducial cross-section limits are up to 30\% stronger at 4~\TeV, rising to 40\% at 5.7~\TeV in the combined dilepton channel using the asymptotic approximation compared to when using pseudo-data. However, the change in the lower mass limit is less than 2\%.

The lower mass limits from the search using Run~3 data are summarised and compared with Run~2 limits in Table~\ref{Tbl:masslimits}. The observed limits in the $ee$ channel are comparable to those expected and exceed those observed in Run~2 for all three $Z^{\prime}$ models. However, for the $\mu\mu$ channel, the limits set are lower than expected for all models, and also lower than the Run~2 limits due to the absence of events above 2.75~\TeV in the Run~2 data sample and the presence of the 3.96~\TeV event in Run~3.

\begin{table}
\begin{center}
\caption{Observed and expected 95\% CL lower mass limits on $Z^{\prime}$ models with varying widths, to the nearest 100~\GeV, in the $ee$, $\mu\mu$ channels and the combined dilepton channel for Run~2, Run~3, and their combination.}
\begin{tabular}{ |c|c|c|c|c|c|c|c| } \hline
\multicolumn{8}{|c|}{Lower mass limits [\TeV]} \\ \hline
\multicolumn{2}{|c|}{Model ($\Gamma/m_{\text{pole}}$)}  &  \multicolumn{2}{c|}{$Z^{\prime}_{\psi}$ (0.5 \%)} &  \multicolumn{2}{c|}{$Z^{\prime}_{\chi}$ (1.2 \%)} &  \multicolumn{2}{c|}{$Z^{\prime}_{\mathrm{SSM}}$ (3.0 \%)} \\ \hline
Channel & Limit Type  & Run~2 & Run~3 & Run~2 & Run~3 & Run~2 & Run~3 \\ \hline
\multirow{2}{*}{$ee$}  & Expected & 4.3 & 4.6 & 4.6 & 4.9 & 4.9 & 5.2 \\
& Observed & 4.1 & 4.6 & 4.6 & 4.8 & 4.9 & 5.1  \\ \hline
\multirow{2}{*}{$\mu\mu$} & Expected & 4.0 & 4.1 & 4.2 & 4.3 & 4.5 & 4.7\\
& Observed &  4.0 & 3.8 & 4.2 & 4.0 & 4.5 & 4.3\\ \hline
\multirow{2}{*}{$ee + \mu\mu$}  & Expected &  4.5 & 4.7 & 4.8 & 4.9 & 5.1 & 5.2 \\
& Observed &  4.5 & 4.6 & 4.8 & 4.9 & 5.1 & 5.2 \\ \hline
\multicolumn{2}{|c|}{} &  \multicolumn{2}{c|}{Run~2 + Run~3} & \multicolumn{2}{c|}{Run~2 + Run~3} & \multicolumn{2}{c|}{Run~2 + Run~3} \\ \hline
\multirow{2}{*}{$ee + \mu\mu$} & Expected & \multicolumn{2}{c|}{4.9} & \multicolumn{2}{c|}{5.1} & \multicolumn{2}{c|}{5.5}\\
& Observed & \multicolumn{2}{c|}{4.8} & \multicolumn{2}{c|}{5.1} & \multicolumn{2}{c|}{5.5}\\ \hline
\end{tabular}
\label{Tbl:masslimits}
\end{center}
\end{table}

\begin{figure}[h]
\centering
\includegraphics[width=0.6\textwidth]{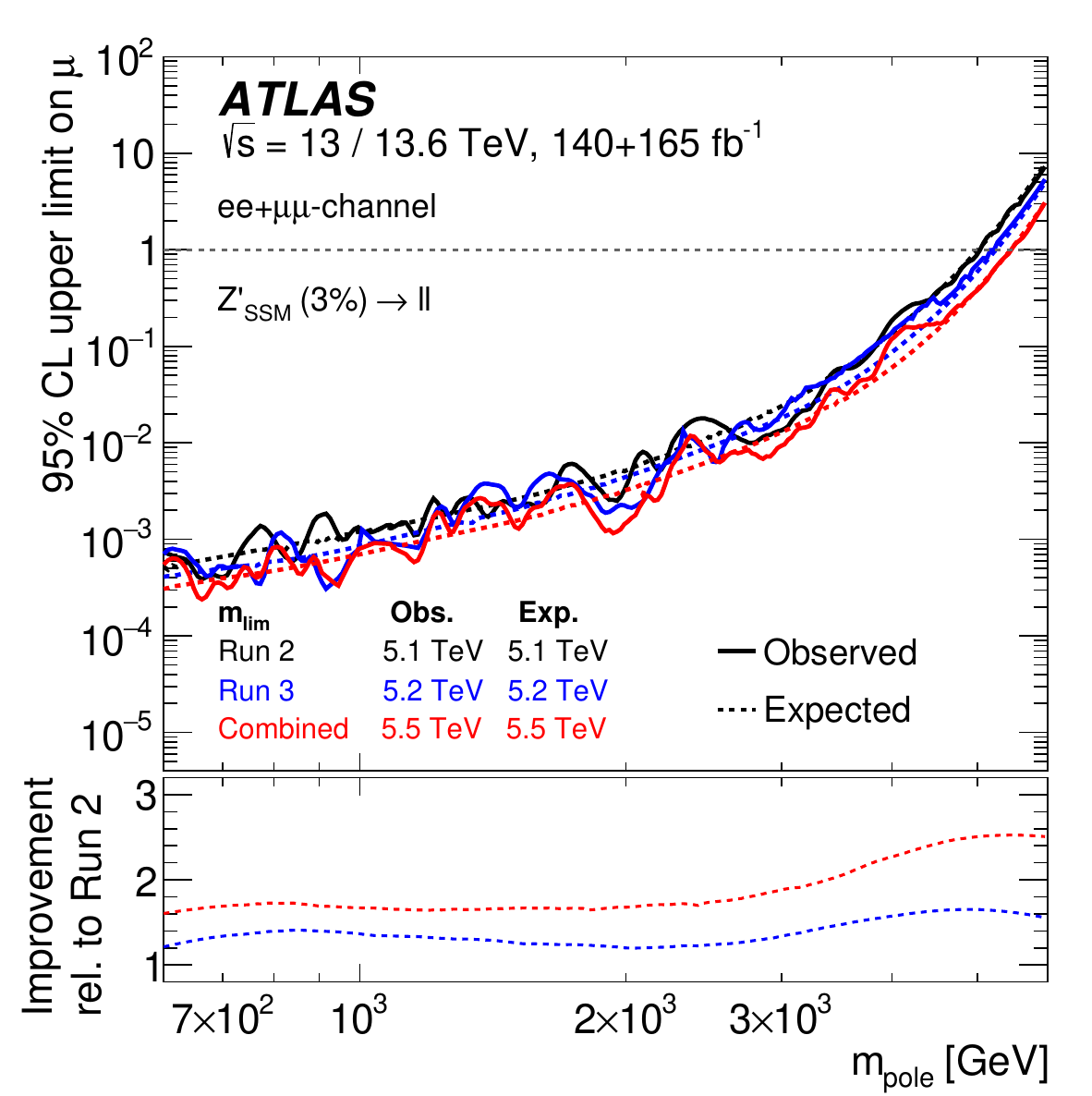}
\caption{Observed and expected upper limits at 95\% CL on the signal strength ($\mu$) as a function of pole mass for the $Z'_{\text{SSM}}$ (3\% relative width) signal showing the combined dilepton channel, for the Run~2, Run~3 and the combined data sample corresponding to a total luminosity of $305~\text{fb}^{-1}$. Observed limits are shown as a solid line and expected limits as a dashed line. The lower panel shows the fractional improvements in the expected limits for Run~3 (blue) and the combined data sample (red) relative to the Run~2 data sample.}
\label{fig:model_limits_combined}
\end{figure}

The combination of the Run~2 and Run~3 $Z'_{\text{SSM}}$ limits is shown in Figure~\ref{fig:model_limits_combined}, compared with the individual Run~2 and Run~3 limits, all in the combined dilepton channel. In the bottom panel of the figure, the comparison of the Run~3 and combined expected limits with those from Run~2 is provided, showing an improvement of between $25\%$ and $50\%$.

\FloatBarrier

\section{Conclusion}

A search is performed for new heavy resonances in the dilepton invariant mass spectrum in the $ee$ and $\mu\mu$ channels, and the combined dilepton channel, using the ATLAS detector at the LHC with 2022--2024 Run~3 $pp$ collision data (165~fb$^{-1}$ at $\sqrt{s} =$ 13.6~\TeV). These results are statistically combined with Run~2 results (140~fb$^{-1}$ at $\sqrt{s} =$ 13~\TeV). A functional form is used to fit the dilepton invariant mass spectra in data, describing the background shape and having the flexibility to account for generic signal shapes of various widths and pole masses. No significant deviations from the SM are observed globally. Therefore, limits are set on the fiducial cross-section times branching ratio in the dilepton channel, and mass limits extracted for three benchmark models, of 4.8 (4.9), 5.1 (5.1), and 5.5 (5.5)~\TeV observed (expected) for the $Z^{\prime}_{\mathrm{\psi}}$, $Z^{\prime}_{\mathrm{\chi}}$ and $Z^{\prime}_{\mathrm{SSM}}$ boson, respectively, representing the most stringent limits to date on high-mass dilepton resonances on the considered $Z'$ models.

\newpage

\section*{Acknowledgements}
%

%

% The next lines are included from the .//acknowledgements/Acknowledgements.tex input file
%
%

%
%

We thank CERN for the very successful operation of the LHC and its injectors, as well as the support staff at
CERN and at our institutions worldwide without whom ATLAS could not be operated efficiently.

The crucial computing support from all WLCG partners is acknowledged gratefully, in particular from CERN, the ATLAS Tier-1 facilities at TRIUMF/SFU (Canada), NDGF (Denmark, Norway, Sweden), CC-IN2P3 (France), KIT/GridKA (Germany), INFN-CNAF (Italy), NL-T1 (Netherlands), PIC (Spain), RAL (UK) and BNL (USA), the Tier-2 facilities worldwide and large non-WLCG resource providers. Major contributors of computing resources are listed in Ref.~\cite{ATL-SOFT-PUB-2026-001}.

We gratefully acknowledge the support of ANPCyT, Argentina; YerPhI, Armenia; ARC, Australia; BMWFW and FWF, Austria; ANAS, Azerbaijan; CNPq and FAPESP, Brazil; NSERC, NRC and CFI, Canada; CERN; ANID, Chile; CAS, MOST, NSFC and FRFCU, China; Minciencias, Colombia; MEYS CR, Czech Republic; DNRF and DNSRC, Denmark; IN2P3-CNRS and CEA-DRF/IRFU, France; SRNSFG, Georgia; BMFTR, HGF and MPG, Germany; GSRI, Greece; RGC and Hong Kong SAR, China; ICHEP and Academy of Sciences and Humanities, Israel; INFN, Italy; MEXT and JSPS, Japan; CNRST, Morocco; NWO, Netherlands; RCN, Norway; MNiSW, Poland; FCT, Portugal; MNE/IFA, Romania; MSTDI, Serbia; MSSR, Slovakia; ARIS and MVZI, Slovenia; DSI/NRF, South Africa; MICIU/AEI, Spain; SRC and Wallenberg Foundation, Sweden; SERI, SNSF and Cantons of Bern and Geneva, Switzerland; NSTC, Taipei; TENMAK, T\"urkiye; STFC/UKRI, United Kingdom; DOE and NSF, United States of America.

Individual groups and members have received support from BCKDF, CANARIE, CRC and DRAC, Canada; CERN-CZ and PRIMUS, Czech Republic; COST, ERC, ERDF, Horizon 2020 and Marie Sk{\l}odowska-Curie Actions, European Union; Investissements d'Avenir Labex, Investissements d'Avenir Idex and ANR, France; DFG and AvH Foundation, Germany; Herakleitos, Thales and Aristeia programmes co-financed by EU-ESF and the Greek NSRF, Greece; BSF-NSF and MINERVA, Israel; NCN and NAWA, Poland; La Caixa Banking Foundation, CERCA and AGAUR programs from Generalitat de Catalunya and PROMETEO and GenT Programmes Generalitat Valenciana, Spain; G\"{o}ran Gustafssons Stiftelse, Sweden; The Royal Society and Leverhulme Trust, United Kingdom; Eric and Wendy Schmidt Fund for Strategic Innovation, United States of America.

In addition, individual members wish to acknowledge support from Chile: Agencia Nacional de Investigaci\'on y Desarrollo (ANID FONDECYT reg. 1230987, FONDECYT 1230812, FONDECYT 1240864, Fondecyt 3240661, Fondecyt Regular 1240721); China: Fundamental Research Funds for the Central Universities (010-63263105), Chinese Ministry of Science and Technology (MOST-2023YFA1605700, MOST-2023YFA1609300), National Natural Science Foundation of China (NSFC 12275265, NSFC-W2543005); Czech Republic: Czech Science Foundation (GACR - 24-11373S), Ministry of Education Youth and Sports (ERC-CZ-LL2327, FORTE CZ.02.01.01/00/22\_008/0004632); EU: H2020 European Research Council (ERC - 101002463); European Union: European Research Council (BARD No. 101116429, ERC - 101219398, ERC 101089007), European Regional Development Fund (HE COFUND GA No.101081355, ERDF), Marie Sklodowska-Curie Actions (GAP-101168829); France: Agence Nationale de la Recherche (ANR-22-EDIR-0002, ANR-24-CE31-0504-01); Germany: Deutsche Forschungsgemeinschaft (DFG - 469666862); China: Research Grants Council (GRF); Italy: Istituto Nazionale di Fisica Nucleare (LHC-MIUR - 28003/2025), Ministero dell'Università e della Ricerca (NextGenEU I53D23000820006 M4C2.1.1, SOE2024\_0000023); Japan: Japan Society for the Promotion of Science (JSPS KAKENHI  JP25H0063, JSPS KAKENHI JP22H04944, JSPS KAKENHI JP24K23939, JSPS KAKENHI JP24KK0251, JSPS KAKENHI JP25H00650, JSPS KAKENHI JP25H01291, JSPS KAKENHI JP25K01011, JSPS KAKENHI JP25K01023, JSPS KAKENHI JP25KK0047); Poland: Polish National Science Centre (NCN 2021/42/E/ST2/00350, NCN OPUS 2023/51/B/ST2/02507, NCN OPUS nr 2022/47/B/ST2/03059, NCN UMO-2019/34/E/ST2/00393, UMO-2022/47/O/ST2/00148, UMO-2023/49/B/ST2/04085, UMO-2023/51/B/ST2/00920, UMO-2024/53/N/ST2/00869); Spain: Agència de Gestió d'Ajuts Universitaris i de Recerca. (AGAUR - 2023 BP 00141), Ministry of Science and Innovation (RYC2019-028510-I, RYC2020-030254-I, RYC2021-031273-I, RYC2022-038164-I), Ministerio de Ciencia, Innovación y Universidades/Agencia Estatal de Investigaci\'on (EU NextGenerationEU (PRTR-C17.I1), PID2022-142604OB-C22); Sweden: Carl Trygger Foundation (Carl Trygger Foundation CTS 22:2312), Swedish Research Council (Swedish Research Council 2023-04654, VR 2021-03651, VR 2022-03845, VR 2022-04683, VR 2023-03403, VR 2024-05451, VR 2025-05940), Knut and Alice Wallenberg Foundation (KAW 2023.0366); United Kingdom: The Binks Trust, Royal Society (NIF-R1-231091); United States of America: U.S. Department of Energy (ECA DE-AC02-76SF00515), John Templeton Foundation (John Templeton Foundation 63206), Neubauer Family Foundation.

%
%

% End of text imported from the .//acknowledgements/Acknowledgements.tex input file

%
%
%
%
%
%
%
%
%

%
\clearpage

\clearpage

\printbibliography

@article{JPsiDiscovery,
    title = {Discovery of a Narrow Resonance in ${e}^{+}{e}^{\ensuremath{-}}$ Annihilation},
    author = {Augustin, J.-E. and Boyarski, A. M. and Breidenbach, M. and Bulos, F. and Dakin, J. T. and Feldman, G. J. and Fischer, G. E. and Fryberger, D. and Hanson, G. and Jean-Marie, B. and Larsen, R. R. and L\"uth, V. and Lynch, H. L. and Lyon, D. and Morehouse, C. C. and Paterson, J. M. and Perl, M. L. and Richter, B. and Rapidis, P. and Schwitters, R. F. and Tanenbaum, W. M. and Vannucci, F. and Abrams, G. S. and Briggs, D. and Chinowsky, W. and Friedberg, C. E. and Goldhaber, G. and Hollebeek, R. J. and Kadyk, J. A. and Lulu, B. and Pierre, F. and Trilling, G. H. and Whitaker, J. S. and Wiss, J. and Zipse, J. E.},
    journal = {Phys. Rev. Lett.},
    volume = {33},
    issue = {23},
    pages = {1406--1408},
    numpages = {0},
    year = {1974},
    month = {12},
    publisher = {American Physical Society},
    doi = {10.1103/PhysRevLett.33.1406},
}

@article{JPsiDiscovery2,
    title = {The discovery of the $J$ particle: A personal recollection},
    author = {Samuel C. C. Ting},
    journal = {Rev. Mod. Phys.},
    volume = {49},
    issue = {235},
    year = {1977},
    month = {04},
}

@article{UpsilonDiscovery,
    title = {Observation of a Dimuon Resonance at 9.5 GeV in 400-GeV Proton-Nucleus Collisions},
    author = {Herb, S. W. and Hom, D. C. and Lederman, L. M. and Sens, J. C. and Snyder, H. D. and Yoh, J. K. and Appel, J. A. and Brown, B. C. and Brown, C. N. and Innes, W. R. and Ueno, K. and Yamanouchi, T. and Ito, A. S. and J\"ostlein, H. and Kaplan, D. M. and Kephart, R. D.},
    journal = {Phys. Rev. Lett.},
    volume = {39},
    issue = {5},
    pages = {252--255},
    numpages = {0},
    year = {1977},
    month = {08},
    publisher = {American Physical Society},
    doi = {10.1103/PhysRevLett.39.252},
}

@article{SMZDiscovery,
    title = {Experimental observation of lepton pairs of invariant mass around 95 GeV/$c^{2}$ at the CERN SPS collider},
    author = {{UA1 Collaboration}},
    journal = { Phys. Lett. B},
    volume = {126},
    number = {5},
    pages = {398-410},
    year = {1983},
    issn = {0370-2693},
    doi = {https://doi.org/10.1016/0370-2693(83)90188-0},
}

@article{BW1,
    author = "Kycia, Rados{\l}aw A. and Jadach, Stanis{\l}aw",
    title = "{Relativistic Voigt profile for unstable particles in high energy physics}",
    eprint = "1711.09304",
    archivePrefix = "arXiv",
    primaryClass = "math-ph",
    doi = "10.1016/j.jmaa.2018.03.065",
    journal = "J. Math. Anal. Appl.",
    volume = "463",
    number = "2",
    pages = "1040--1051",
    year = "2018"
}

@article{BW2,
    author = "Matthews, P. T. and Salam, Abdus",
    title = "{Relativistic field theory of unstable particles}",
    doi = "10.1103/PhysRev.112.283",
    journal = "Phys. Rev.",
    volume = "112",
    pages = "283--287",
    year = "1958"
}

@article{BAGNAIA1983130,
    author = {{UA2 Collaboration}},
    title = "{Evidence for $Z^{0} \to e^+ e^-$ at the CERN $\bar{p} p$ Collider}",
    reportNumber = "CERN-EP-83-112",
    doi = "10.1016/0370-2693(83)90744-X",
    journal = "Phys. Lett. B",
    volume = "129",
    pages = "130--140",
    year = "1983"
}

@article{GUTZPrimes,
    title = {New heavy gauge bosons in pp and $p\bar{p}$ collisions},
    author = {Langacker, Paul and Robinett, Richard W. and Rosner, Jonathan L.},
    journal = {Phys. Rev. D},
    volume = {30},
    issue = {7},
    pages = {1470--1487},
    year = {1984},
    publisher = {American Physical Society},
    doi = {10.1103/PhysRevD.30.1470}
}

@article{E6ZPrimes,
    title = {Extra gauge bosons in ${\mathrm{E}}_{6}$},
    author = {London, David and Rosner, Jonathan L.},
    journal = {Phys. Rev. D},
    volume = {34},
    issue = {5},
    pages = {1530--1546},
    year = {1986},
    publisher = {American Physical Society},
    doi = {10.1103/PhysRevD.34.1530}
}

@article{Gross_2010,
    author = "Gross, Eilam and Vitells, Ofer",
    title = "{Trial factors for the look elsewhere effect in high energy physics}",
    eprint = "1005.1891",
    archivePrefix = "arXiv",
    primaryClass = "physics.data-an",
    doi = "10.1140/epjc/s10052-010-1470-8",
    journal = "Eur. Phys. J. C",
    volume = "70",
    pages = "525--530",
    year = "2010"
}

@article{Randall:1999ee,
    author = "Randall, Lisa and Sundrum, Raman",
    title = "{Large mass hierarchy from a small extra dimension}",
    eprint = "hep-ph/9905221",
    archivePrefix = "arXiv",
    reportNumber = "MIT-CTP-2860, PUPT-1860, BUHEP-99-9",
    doi = "10.1103/PhysRevLett.83.3370",
    journal = "Phys. Rev. Lett.",
    volume = "83",
    pages = "3370--3373",
    year = "1999"
}

@article{JpsiDiscoveryTing,
  author = {Aubert, J. J. and Becker, U. and Biggs, P. J. and Burger, J. and Chen, M. and Everhart, G. and Goldhagen, P. and Leong, J. and McCorriston, T. and Rhoades, T. G. and Rohde, M. and Ting, Samuel C. C. and Wu, Sau Lan and Lee, Y. Y.},
  title = {Experimental Observation of a Heavy Particle $J$},
  volume = {33},
  issue = {23},
  pages = {1404--1406},
  year = {1974},
  journal = {Phys. Rev. Lett.},
  doi = {10.1103/PhysRevLett.33.1404}
}

@article{Arkani-Hamed:1998jmv,
    author = "Arkani-Hamed, Nima and Dimopoulos, Savas and Dvali, G. R.",
    title = "{The Hierarchy problem and new dimensions at a millimeter}",
    eprint = "hep-ph/9803315",
    archivePrefix = "arXiv",
    reportNumber = "SLAC-PUB-7769, SU-ITP-98-13",
    doi = "10.1016/S0370-2693(98)00466-3",
    journal = "Phys. Lett. B",
    volume = "429",
    pages = "263--272",
    year = "1998"
}

@Article{Arbuzov2016,
    author={Arbuzov, A.
    and Bardin, D.
    and Bondarenko, S.
    and Christova, P.
    and Kalinovskaya, L.
    and Klein, U.
    and Kolesnikov, V.
    and Rumyantsev, L.
    and Sadykov, R.
    and Sapronov, A.},
    title={Update of the MCSANC Monte Carlo integrator, v. 1.20},
    journal={JETP Letters},
    year={2016},
    month={01},
    day={01},
    volume={103},
    number={2},
    pages={131-136},
    issn={1090-6487},
    doi={10.1134/S0021364016020041},
}

@article{dierckx1975algorithm,
      title={An algorithm for smoothing, differentiation and integration of experimental data using spline functions},
      author={Dierckx, Paul},
      journal={Journal of Computational and Applied Mathematics},
      volume={1},
      number={3},
      pages={165--184},
      year={1975},
      publisher={Elsevier}
}

@article{dierckx1982fast,
  title={A fast algorithm for smoothing data on a rectangular grid while using spline functions},
  author={Dierckx, Paul},
  journal={SIAM Journal on Numerical Analysis},
  volume={19},
  number={6},
  pages={1286--1304},
  year={1982},
  publisher={SIAM}
}

@article{dierckx1981improved,
  title={An improved algorithm for curve fitting with spline functions},
  author={Dierckx, Paul},
  journal={TW Reports},
  year={1981},
  publisher={Department of Computer Science, KU Leuven, Leuven, Belgium}
}

@book{dierckx1995curve,
  title={Curve and surface fitting with splines},
  author={Dierckx, Paul},
  year={1995},
  publisher={Oxford University Press}
}

@article{Lai:2010vv,
      author         = "Lai, H.-L. and others",
      title          = "{New parton distributions for collider physics}",
      journal        = "Phys. Rev. D",
      volume         = "82",
      pages          = "074024",
      doi            = "10.1103/PhysRevD.82.074024",
      year           = "2010",
      eprint         = "1007.2241",
      archivePrefix  = "arXiv",
      primaryClass   = "hep-ph",
      reportNumber   = "MSUHEP-100707, SMU-HEP-10-10",
      SLACcitation   = "%%CITATION = ARXIV:1007.2241;%%",
}

@article{Ball:2012cx,
      author         = "{NNPDF Collaboration} and Ball, Richard D. and others",
      title          = "{Parton distributions with LHC data}",
      journal        = "Nucl. Phys. B",
      volume         = "867",
      year           = "2013",
      pages          = "244",
      doi            = "10.1016/j.nuclphysb.2012.10.003",
      eprint         = "1207.1303",
      archivePrefix  = "arXiv",
      primaryClass   = "hep-ph",
      reportNumber   = "EDINBURGH-2012-08, IFUM-FT-997, FR-PHENO-2012-014,
                        RWTH-TTK-12-25, CERN-PH-TH-2012-037, SFB-CPP-12-47\,
                        --CERN-PH-TH-2012-037",
      SLACcitation   = "%%CITATION = ARXIV:1207.1303;%%"
}

@article{Ball:2014uwa,
      author         = "{NNPDF Collaboration} and Ball, Richard D. and others",
      title          = "{Parton distributions for the LHC run II}",
      journal        = "JHEP",
      volume         = "04",
      year           = "2015",
      pages          = "040",
      doi            = "10.1007/JHEP04(2015)040",
      eprint         = "1410.8849",
      archivePrefix  = "arXiv",
      primaryClass   = "hep-ph",
      reportNumber   = "EDINBURGH-2014-15, IFUM-1034-FT, CERN-PH-TH-2013-253,
                        OUTP-14-11P, CAVENDISH-HEP-14-11",
      SLACcitation   = "%%CITATION = ARXIV:1410.8849;%%"
}

@article{Sjostrand:2014zea,
      author         = "Sj{\"o}strand, Torbj{\"o}rn and Ask, Stefan and Christiansen,
                        Jesper R. and Corke, Richard and Desai, Nishita and Ilten,
                        Philip and Mrenna, Stephen and Prestel, Stefan and
                        Rasmussen, Christine O. and Skands, Peter Z.",
      title          = "{An introduction to PYTHIA 8.2}",
      journal        = "Comput. Phys. Commun.",
      volume         = "191",
      year           = "2015",
      pages          = "159",
      doi            = "10.1016/j.cpc.2015.01.024",
      eprint         = "1410.3012",
      archivePrefix  = "arXiv",
      primaryClass   = "hep-ph",
      reportNumber   = "LU-TP-14-36, MCNET-14-22, CERN-PH-TH-2014-190,
                        FERMILAB-PUB-14-316-CD, DESY-14-178, SLAC-PUB-16122,
                        --FERMILAB-PUB-14-316-CD",
      SLACcitation   = "%%CITATION = ARXIV:1410.3012;%%"
}

@Article{Nason:2004rx,
      author         = "Nason, Paolo",
      title          = "{A new method for combining NLO QCD with shower Monte Carlo algorithms}",
      journal        = "JHEP",
      volume         = "11",
      pages          = "040",
      doi            = "10.1088/1126-6708/2004/11/040",
      year           = "2004",
      eprint         = "hep-ph/0409146",
      archivePrefix  = "arXiv",
}

@Article{Frixione:2007vw,
      author         = "Frixione, Stefano and Nason, Paolo and Oleari, Carlo",
      title          = "{Matching NLO QCD computations with parton shower
                        simulations: the POWHEG method}",
      journal        = "JHEP",
      volume         = "11",
      pages          = "070",
      doi            = "10.1088/1126-6708/2007/11/070",
      year           = "2007",
      eprint         = "0709.2092",
      archivePrefix  = "arXiv",
      primaryClass   = "hep-ph",
}

@Article{Alioli:2010xd,
      author         = "Alioli, Simone and Nason, Paolo and Oleari, Carlo and Re,
                        Emanuele",
      title          = "{A general framework for implementing NLO calculations in
                        shower Monte Carlo programs: the POWHEG BOX}",
      journal        = "JHEP",
      volume         = "06",
      pages          = "043",
      doi            = "10.1007/JHEP06(2010)043",
      year           = "2010",
      eprint         = "1002.2581",
      archivePrefix  = "arXiv",
      primaryClass   = "hep-ph",
}

@article{Anastasiou:2003ds,
      author         = "Anastasiou, Charalampos and Dixon, Lance and Melnikov,
                        Kirill and Petriello, Frank",
      title          = "{High-precision QCD at hadron colliders: Electroweak
                        gauge boson rapidity distributions at next-to-next-to leading order}",
      journal        = "Phys. Rev. D",
      volume         = "69",
      year           = "2004",
      pages          = "094008",
      doi            = "10.1103/PhysRevD.69.094008",
      eprint         = "hep-ph/0312266",
      archivePrefix  = "arXiv",
      reportNumber   = "SLAC-PUB-10288, UH-511-1042-03",
      SLACcitation   = "%%CITATION = HEP-PH/0312266;%%"
}

@article{Frixione:2007nw,
      author         = "Frixione, Stefano and Ridolfi, Giovanni and Nason, Paolo",
      title          = "{A positive-weight next-to-leading-order Monte Carlo for
                        heavy flavour hadroproduction}",
      journal        = "JHEP",
      volume         = "09",
      pages          = "126",
      doi            = "10.1088/1126-6708/2007/09/126",
      year           = "2007",
      eprint         = "0707.3088",
      archivePrefix  = "arXiv",
      primaryClass   = "hep-ph",
      SLACcitation   = "%%CITATION = ARXIV:0707.3088;%%",
}

@article{Bothmann:2019yzt,
      author = "Bothmann, Enrico and others",
      title = "{Event generation with Sherpa 2.2}",
      journal = "SciPost Phys.",
      volume = "7",
      year = "2019",
      number = "3",
      pages = "034",
      doi = "10.21468/SciPostPhys.7.3.034",
      reportNumber = "FERMILAB-PUB-19-218-T, SLAC-PUB-17433, IPPP/19/42, MCNET-19-11",
      eprint = "1905.09127",
      archivePrefix = "arXiv",
      primaryClass = "hep-ph",
}

@article{Agostinelli:2002hh,
      author         = "Agostinelli, S. and others",
      title          = "{\textsc{Geant4} -- a simulation toolkit}",
      journal        = "Nucl. Instrum. Meth. A",
      volume         = "506",
      year           = "2003",
      pages          = "250",
      doi            = "10.1016/S0168-9002(03)01368-8",
      reportNumber   = "SLAC-PUB-9350, FERMILAB-PUB-03-339",
      SLACcitation   = "%%CITATION = NUIMA,A506,250;%%"
}

@Article{Read:2002hq,
    Author = {Read, Alexander L.},
    Journal = {J. Phys. G},
    Pages = {2693},
    doi = "10.1088/0954-3899/28/10/313",
    Title = "{Presentation of search results: the \(CL_s\) technique}",
    Volume = {28},
    Year = {2002}
}

@Article{Cowan:2010js,
   Author = {Cowan, Glen and Cranmer, Kyle and Gross, Eilam and Vitells, Ofer},
   Title = "{Asymptotic formulae for likelihood-based tests of new physics}",
   Journal = {Eur. Phys. J. C},
   Volume = {71},
   Year = {2011},
   Pages = {1554},
   doi = "10.1140/epjc/s10052-011-1554-0",
   Eprint = "1007.1727",
   Archiveprefix = {arXiv},
   Primaryclass = {physics.data-an},
   related = "Cowan:2010js-err",
   relatedstring  = "Erratum:",
}

@article{Zyla:2020zbs,
    author = "{Particle Data Group} and Zyla, P.A. and others",
    title = "{Review of Particle Physics}",
    journal = "Prog. Theor. Exp. Phys.",
    volume = "2020",
    number = "8",
    pages = "083C01",
    year = "2020",
    doi = "10.1093/ptep/ptaa104",
}

@Article{PERF-2007-01,
    author         = "{ATLAS Collaboration}",
    title          = "{The ATLAS Experiment at the CERN Large Hadron Collider}",
    journal        = "JINST",
    volume         = "3",
    year           = "2008",
    pages          = "S08003",
    doi            = "10.1088/1748-0221/3/08/S08003",
    primaryClass   = "hep-ex",
}

@Article{EXOT-2010-09,
    author         = "{ATLAS Collaboration}",
    title          = "{Search for high mass dilepton resonances in \(pp\) collisions at \(\sqrt{s} = 7\,\text{TeV}\) with the ATLAS experiment}",
    journal        = "Phys. Lett. B",
    volume         = "700",
    year           = "2011",
    pages          = "163",
    doi            = "10.1016/j.physletb.2011.04.044",
    eprint         = "1103.6218",
    archivePrefix  = "arXiv",
    primaryClass   = "hep-ex",
}

@Article{EXOT-2011-06,
    author         = "{ATLAS Collaboration}",
    title          = "{Search for Dilepton Resonances in \(pp\) Collisions at \(\sqrt{s} = 7\,\text{TeV}\) with the ATLAS Detector}",
    journal        = "Phys. Rev. Lett.",
    volume         = "107",
    year           = "2011",
    pages          = "272002",
    doi            = "10.1103/PhysRevLett.107.272002",
    reportNumber   = "CERN-PH-EP-2011-123",
    eprint         = "1108.1582",
    archivePrefix  = "arXiv",
    primaryClass   = "hep-ex",
}

@Article{EXOT-2012-01,
    author         = "{ATLAS Collaboration}",
    title          = "{Search for high-mass resonances decaying to dilepton final states in \(pp\) collisions at \(\sqrt{s} = 7\,\text{TeV}\) with the ATLAS detector}",
    journal        = "JHEP",
    volume         = "11",
    year           = "2012",
    pages          = "138",
    doi            = "10.1007/JHEP11(2012)138",
    reportNumber   = "CERN-PH-EP-2012-190",
    eprint         = "1209.2535",
    archivePrefix  = "arXiv",
    primaryClass   = "hep-ex",
}

@Article{EXOT-2012-23,
    author         = "{ATLAS Collaboration}",
    title          = "{Search for high-mass dilepton resonances in \(pp\) collisions at \(\sqrt{s} = 8\,\text{TeV}\) with the ATLAS detector}",
    journal        = "Phys. Rev. D",
    volume         = "90",
    year           = "2014",
    pages          = "052005",
    doi            = "10.1103/PhysRevD.90.052005",
    reportNumber   = "CERN-PH-EP-2014-053",
    eprint         = "1405.4123",
    archivePrefix  = "arXiv",
    primaryClass   = "hep-ex",
}

@Article{STDM-2012-23,
    author         = "{ATLAS Collaboration}",
    title          = "{Measurement of the \(Z/\gamma^*\) boson transverse momentum distribution in \(pp\) collisions at \(\sqrt{s} = 7\,\text{TeV}\) with the ATLAS detector}",
    journal        = "JHEP",
    volume         = "09",
    year           = "2014",
    pages          = "145",
    doi            = "10.1007/JHEP09(2014)145",
    reportNumber   = "CERN-PH-EP-2014-075",
    eprint         = "1406.3660",
    archivePrefix  = "arXiv",
    primaryClass   = "hep-ex",
}

@Article{EXOT-2015-07,
    author         = "{ATLAS Collaboration}",
    title          = "{Search for high-mass new phenomena in the dilepton final state using proton--proton collisions at \(\sqrt{s} = 13\,\text{TeV}\) with the ATLAS detector}",
    journal        = "Phys. Lett. B",
    volume         = "761",
    year           = "2016",
    pages          = "372",
    doi            = "10.1016/j.physletb.2016.08.055",
    reportNumber   = "CERN-EP-2016-116",
    eprint         = "1607.03669",
    archivePrefix  = "arXiv",
    primaryClass   = "hep-ex",
}

@Article{EXOT-2016-05,
    author         = "{ATLAS Collaboration}",
    title          = "{Search for new high-mass phenomena in the dilepton final state using \(36\,\text{fb}^{-1}\) of proton--proton collision data at \(\sqrt{s} = 13\,\text{TeV}\) with the ATLAS detector}",
    journal        = "JHEP",
    volume         = "10",
    year           = "2017",
    pages          = "182",
    doi            = "10.1007/JHEP10(2017)182",
    reportNumber   = "CERN-EP-2017-119",
    eprint         = "1707.02424",
    archivePrefix  = "arXiv",
    primaryClass   = "hep-ex",
}

@Article{TRIG-2016-01,
    author         = "{ATLAS Collaboration}",
    title          = "{Performance of the ATLAS trigger system in 2015}",
    journal        = "Eur. Phys. J. C",
    volume         = "77",
    year           = "2017",
    pages          = "317",
    doi            = "10.1140/epjc/s10052-017-4852-3",
    reportNumber   = "CERN-EP-2016-241",
    eprint         = "1611.09661",
    archivePrefix  = "arXiv",
    primaryClass   = "hep-ex",
}

@Article{PERF-2017-01,
    author         = "{ATLAS Collaboration}",
    title          = "{Electron reconstruction and identification in the ATLAS experiment using the 2015 and 2016 LHC proton--proton collision data at \(\sqrt{s} = 13\,\text{TeV}\)}",
    journal        = "Eur. Phys. J. C",
    volume         = "79",
    year           = "2019",
    pages          = "639",
    doi            = "10.1140/epjc/s10052-019-7140-6",
    reportNumber   = "CERN-EP-2018-273",
    eprint         = "1902.04655",
    archivePrefix  = "arXiv",
    primaryClass   = "physics.ins-det",
}

@Article{PERF-2017-03,
    author         = "{ATLAS Collaboration}",
    title          = "{Electron and photon energy calibration with the ATLAS detector using 2015--2016 LHC proton--proton collision data}",
    journal        = "JINST",
    volume         = "14",
    year           = "2019",
    pages          = "P03017",
    doi            = "10.1088/1748-0221/14/03/P03017",
    reportNumber   = "CERN-EP-2018-296",
    eprint         = "1812.03848",
    archivePrefix  = "arXiv",
    primaryClass   = "hep-ex",
}

@Article{EGAM-2018-01,
    author         = "{ATLAS Collaboration}",
    title          = "{Electron and photon performance measurements with the ATLAS detector using the 2015--2017 LHC proton--proton collision data}",
    journal        = "JINST",
    volume         = "14",
    year           = "2019",
    pages          = "P12006",
    doi            = "10.1088/1748-0221/14/12/P12006",
    reportNumber   = "CERN-EP-2019-145",
    eprint         = "1908.00005",
    archivePrefix  = "arXiv",
    primaryClass   = "hep-ex",
}

@Article{EXOT-2018-08,
    author         = "{ATLAS Collaboration}",
    title          = "{Search for high-mass dilepton resonances using \(139\,\text{fb}^{-1}\) of \(pp\) collision data collected at \(\sqrt{s} = 13\,\text{TeV}\) with the ATLAS detector}",
    journal        = "Phys. Lett. B",
    volume         = "796",
    year           = "2019",
    pages          = "68",
    doi            = "10.1016/j.physletb.2019.07.016",
    reportNumber   = "CERN-EP-2019-030",
    eprint         = "1903.06248",
    archivePrefix  = "arXiv",
    primaryClass   = "hep-ex",
}

@Article{MUON-2018-03,
    author         = "{ATLAS Collaboration}",
    title          = "{Muon reconstruction and identification efficiency in ATLAS using the full Run~2 \(pp\) collision data set at \(\sqrt{s} = 13\,\text{TeV}\)}",
    journal        = "Eur. Phys. J. C",
    volume         = "81",
    year           = "2021",
    pages          = "578",
    doi            = "10.1140/epjc/s10052-021-09233-2",
    reportNumber   = "CERN-EP-2020-199",
    eprint         = "2012.00578",
    archivePrefix  = "arXiv",
    primaryClass   = "hep-ex",
}

@Article{GENR-2019-02,
    author         = "{ATLAS Collaboration}",
    title          = "{The ATLAS experiment at the CERN Large Hadron Collider: a description of the detector configuration for Run~3}",
    journal        = "JINST",
    volume         = "19",
    year           = "2024",
    pages          = "P05063",
    doi            = "10.1088/1748-0221/19/05/P05063",
    reportNumber   = "CERN-EP-2022-259",
    eprint         = "2305.16623",
    archivePrefix  = "arXiv",
    primaryClass   = "physics.ins-det",
}

@Article{DAPR-2021-01,
    author         = "{ATLAS Collaboration}",
    title          = "{Luminosity determination in \(pp\) collisions at \(\sqrt{s} = 13\,\text{TeV}\) using the ATLAS detector at the LHC}",
    journal        = "Eur. Phys. J. C",
    volume         = "83",
    year           = "2023",
    pages          = "982",
    doi            = "10.1140/epjc/s10052-023-11747-w",
    reportNumber   = "CERN-EP-2022-281",
    eprint         = "2212.09379",
    archivePrefix  = "arXiv",
    primaryClass   = "hep-ex",
}

@Article{EGAM-2021-01,
    author         = "{ATLAS Collaboration}",
    title          = "{Electron and photon efficiencies in LHC Run~2 with the ATLAS experiment}",
    journal        = "JHEP",
    volume         = "05",
    year           = "2024",
    pages          = "162",
    doi            = "10.1007/JHEP05(2024)162",
    reportNumber   = "CERN-EP-2023-182",
    eprint         = "2308.13362",
    archivePrefix  = "arXiv",
    primaryClass   = "hep-ex",
}

@Article{EGAM-2021-02,
    author         = "{ATLAS Collaboration}",
    title          = "{Electron and photon energy calibration with the ATLAS detector using LHC Run~2 data}",
    journal        = "JINST",
    volume         = "19",
    year           = "2024",
    pages          = "P02009",
    doi            = "10.1088/1748-0221/19/02/P02009",
    reportNumber   = "CERN-EP-2023-128",
    eprint         = "2309.05471",
    archivePrefix  = "arXiv",
    primaryClass   = "hep-ex",
}

@Article{MUON-2022-01,
    author         = "{ATLAS Collaboration}",
    title          = "{Studies of the muon momentum calibration and performance of the ATLAS detector with \(pp\) collisions at \(\sqrt{s} = 13\,\text{TeV}\)}",
    journal        = "Eur. Phys. J. C",
    volume         = "83",
    year           = "2023",
    pages          = "686",
    doi            = "10.1140/epjc/s10052-023-11584-x",
    reportNumber   = "CERN-EP-2022-274",
    eprint         = "2212.07338",
    archivePrefix  = "arXiv",
    primaryClass   = "hep-ex",
}

@Article{SOFT-2022-02,
    author         = "{ATLAS Collaboration}",
    title          = "{Software and computing for Run~3 of the ATLAS experiment at the LHC}",
    journal        = "Eur. Phys. J. C",
    volume         = "85",
    year           = "2025",
    pages          = "234",
    doi            = "10.1140/epjc/s10052-024-13701-w",
    reportNumber   = "CERN-EP-2024-100",
    eprint         = "2404.06335",
    archivePrefix  = "arXiv",
    primaryClass   = "hep-ex",
    related        = "SOFT-2022-02-err",
    relatedstring  = "Erratum:",
}

@Article{TRIG-2022-01,
    author         = "{ATLAS Collaboration}",
    title          = "{The ATLAS trigger system for LHC Run~3 and trigger performance in 2022}",
    journal        = "JINST",
    volume         = "19",
    year           = "2024",
    pages          = "P06029",
    doi            = "10.1088/1748-0221/19/06/P06029",
    reportNumber   = "CERN-EP-2023-299",
    eprint         = "2401.06630",
    archivePrefix  = "arXiv",
    primaryClass   = "hep-ex",
}

@Booklet{ATL-SOFT-PUB-2026-001,
    author         = "{ATLAS Collaboration}",
    title          = "{ATLAS Computing Acknowledgements}",
    howpublished   = "{ATL-SOFT-PUB-2026-001}",
    url            = "https://cds.cern.ch/record/2952666",
    year           = "2026",
}

@Article{LUCID2,
    author         = "Avoni, G. and others",
    title          = "{The new LUCID-2 detector for luminosity measurement and monitoring in ATLAS}",
    journal        = "JINST",
    volume         = "13",
    year           = "2018",
    pages          = "P07017",
    doi            = "10.1088/1748-0221/13/07/P07017",
}

@Article{CMS-EXO-19-019,
    author         = "{CMS Collaboration}",
    title          = "{Search for resonant and nonresonant new phenomena in high-mass dilepton final states at \(\sqrt{s} = 13\,\text{TeV}\)}",
    journal        = "JHEP",
    volume         = "07",
    year           = "2021",
    pages          = "208",
    doi            = "10.1007/JHEP07(2021)208",
    reportNumber   = "CERN-EP-2021-026",
    eprint         = "2103.02708",
    archivePrefix  = "arXiv",
    primaryClass   = "hep-ex",
}

@Booklet{ATL-PHYS-PUB-2014-021,
    author         = "{ATLAS Collaboration}",
    title          = "{ATLAS Pythia~8 tunes to \(7\,\text{TeV}\) data}",
    howpublished   = "{ATL-PHYS-PUB-2014-021}",
    url            = "https://cds.cern.ch/record/1966419",
    year           = "2014",
}

@Booklet{ATL-PHYS-PUB-2020-028,
    author         = "{ATLAS Collaboration}",
    title          = "{Recommendations for the Modeling of Smooth Backgrounds}",
    howpublished   = "{ATL-PHYS-PUB-2020-028}",
    url            = "https://cds.cern.ch/record/2743717",
    year           = "2020",
}

@Booklet{ATL-DAPR-PUB-2025-001,
    author         = "{ATLAS Collaboration}",
    title          = "{Preliminary luminosity calibration of the ATLAS \(13.6\,\text{TeV}\) data recorded in 2024 and combination with the 2022 and 2023 measurements}",
    howpublished   = "{ATL-DAPR-PUB-2025-001}",
    url            = "https://cds.cern.ch/record/2948582",
    year           = "2025",
}
\clearpage
% ATLAS Collaboration author list
% Reference date of EXOT-2025-09 is 2026-07-02
% Author list last updated on date 30-JUL-26
% Data extracted on 30-Jul-2026 for paper reference EXOT-2025-09
% at 3:55pm
 
\begin{flushleft}
\hypersetup{urlcolor=black}
{\Large The ATLAS Collaboration}

\bigskip

\AtlasOrcid[0000-0002-6665-4934]{G.~Aad}$^\textrm{\scriptsize 102}$,
\AtlasOrcid[0000-0001-7616-1554]{E.~Aakvaag}$^\textrm{\scriptsize 17}$,
\AtlasOrcid[0000-0002-5888-2734]{B.~Abbott}$^\textrm{\scriptsize 121}$,
\AtlasOrcid[0000-0002-0287-5869]{S.~Abdelhameed}$^\textrm{\scriptsize 83b}$,
\AtlasOrcid[0000-0002-1002-1652]{K.~Abeling}$^\textrm{\scriptsize 54}$,
\AtlasOrcid[0000-0001-5763-2760]{N.J.~Abicht}$^\textrm{\scriptsize 48}$,
\AtlasOrcid[0000-0002-8496-9294]{S.H.~Abidi}$^\textrm{\scriptsize 30}$,
\AtlasOrcid[0009-0003-6578-220X]{M.~Aboelela}$^\textrm{\scriptsize 44}$,
\AtlasOrcid[0000-0002-9987-2292]{A.~Aboulhorma}$^\textrm{\scriptsize 36e}$,
\AtlasOrcid[0000-0001-5329-6640]{H.~Abramowicz}$^\textrm{\scriptsize 154}$
% ...
\bigskip
\\

% ...
$^{17}$Department for Physics and Technology, University of Bergen, Bergen; Norway.\\
% ...
$^{30}$Physics Department, Brookhaven National Laboratory, Upton NY; United States of America.\\
% ...
$^{36}$$^{(a)}$Facult\'e des Sciences Ain Chock, Universit\'e Hassan II de Casablanca;$^{(b)}$Facult\'{e} des Sciences, Universit\'{e} Ibn-Tofail, K\'{e}nitra;$^{(c)}$Facult\'e des Sciences Semlalia, Universit\'e Cadi Ayyad, LPHEA-Marrakech;$^{(d)}$LPMR, Facult\'e des Sciences, Universit\'e Mohamed Premier, Oujda;$^{(e)}$Facult\'e des sciences, Universit\'e Mohammed V, Rabat;$^{(f)}$Institute of Applied Physics, Mohammed VI Polytechnic University, Ben Guerir; Morocco.\\
% ...
$^{44}$Physics Department, Southern Methodist University, Dallas TX; United States of America.\\
% ...
$^{48}$Fakult\"{a}t Physik, Technische Universit{\"a}t Dortmund, Dortmund; Germany.\\
% ...
$^{54}$II. Physikalisches Institut, Georg-August-Universit\"{a}t G\"ottingen, G\"ottingen; Germany.\\
% ...
$^{83}$$^{(a)}$Khalifa University of Science and Technology, Abu Dhabi;$^{(b)}$New York University Abu Dhabi, Abu Dhabi;$^{(c)}$United Arab Emirates University, Al Ain;$^{(d)}$University of Sharjah, Sharjah; United Arab Emirates.\\
% ...
$^{102}$CPPM, Aix-Marseille Universit\'e, CNRS/IN2P3, Marseille; France.\\
% ...
$^{121}$Homer L. Dodge Department of Physics and Astronomy, University of Oklahoma, Norman OK; United States of America.\\
% ...
$^{154}$Raymond and Beverly Sackler School of Physics and Astronomy, Tel Aviv University, Tel Aviv; Israel.\\
% ...
$^{a}$ Also at Affiliated with an institute formerly covered by a cooperation agreement with CERN.\\
$^{b}$ Also at An-Najah National University, Nablus; Palestine.\\
$^{c}$ Also at Borough of Manhattan Community College, City University of New York, New York NY; United States of America.\\
$^{d}$ Also at Center for Interdisciplinary Research and Innovation (CIRI-AUTH), Thessaloniki; Greece.\\
$^{e}$ Also at Centre of Physics of the Universities of Minho and Porto (CF-UM-UP); Portugal.\\
$^{f}$ Also at CERN, Geneva; Switzerland.\\
% ...

\end{flushleft}

% Created with Glance <Atlas.Glance@cern.ch>

%

%
%

%

%
%
%
%

\end{document}